\documentclass[10pt,letterpaper]{article}

\usepackage[margin=0.75in]{geometry}
\usepackage[utf8]{inputenc}
\usepackage[title]{appendix}
\usepackage[sort&compress,square,comma,numbers]{natbib}

\usepackage{svg}
\usepackage{graphicx,subfig,float}
\usepackage{tabularx}
\usepackage{amsmath,amsthm,amssymb,commath,amsfonts}
\usepackage{booktabs,multirow}
\usepackage[normalem]{ulem}
\usepackage[colorlinks=true,bookmarksopen=true]{hyperref}
\usepackage{algpseudocode}
\usepackage{algorithm}
\usepackage{setspace}
\usepackage[normalem]{ulem}
\usepackage{todonotes}
\usepackage{lastpage,fancyhdr}
\newcommand{\UMMDPD}{\texttt{\textsubscript{\textit{USER}}MESO 2.5}~}
\newcommand{\UMTWO}{\texttt{\textsubscript{\textit{USER}}MESO 2.0}~}
\newcommand{\USERMESO}{\texttt{\textsubscript{\textit{USER}}MESO}~}

\begin{document}

\begin{flushleft}
{\Large
\textbf\newline{A GPU-accelerated package for simulation of flow in nanoporous source rocks with many-body dissipative particle dynamics$^{\star}$}
}
\newline
\\
Yidong Xia\textsuperscript{a,*,1},
Ansel Blumers\textsuperscript{a,b,1},
Zhen Li\textsuperscript{c,*},
Lixiang Luo\textsuperscript{d},
Yu-Hang Tang\textsuperscript{e},
Joshua Kane\textsuperscript{f},
Hai Huang\textsuperscript{a},
Matthew Andrew\textsuperscript{g},
Milind Deo\textsuperscript{h},
Jan Goral\textsuperscript{h}
\\
\bigskip
\bf\textsuperscript{a} Energy and Environment Science \& Technology, Idaho National Laboratory, Idaho Falls, ID
\\
\bf\textsuperscript{b} Department of Physics, Brown University, Providence, RI
\\
\bf\textsuperscript{c} Division of Applied Mathematics, Brown University, Providence, RI
\\
\bf\textsuperscript{d} Center of Excellence at ORNL, IBM, Oak Ridge, TN
\\
\bf\textsuperscript{e} Computational Research Division, Lawrence Berkeley National Laboratory, Berkeley, CA
\\
\bf\textsuperscript{f} Materials and Fuels Complex, Idaho National Laboratory, Idaho Falls, ID
\\
\bf\textsuperscript{g} Carl Zeiss X-ray Microscopy, Pleasanton, CA
\\
\bf\textsuperscript{h} Department of Chemical Engineering, University of Utah, Salt Lake City, UT
\\
\bigskip
$\star$ Approved for external release:  INL/JOU-19-52933
\\
* Corresponding authors: yidong.xia@inl.gov (Yidong Xia), zhen\_li@brown.edu (Zhen Li)
\\
\bf\textsuperscript{1} These authors contributed equally to the work

\end{flushleft}

\section*{Abstract}
Mesoscopic simulations of hydrocarbon flow in source shales are challenging, in part due to the heterogeneous shale pores with sizes ranging from a few nanometers to a few micrometers. Additionally, the sub-continuum fluid-fluid and fluid-solid interactions in nano- to micro-scale shale pores, which are physically and chemically sophisticated, must be captured. To address those challenges, we present a GPU-accelerated package for simulation of flow in nano- to micro-pore networks with a many-body dissipative particle dynamics (mDPD) mesoscale model. Based on a fully distributed parallel paradigm, the code offloads all intensive workloads on GPUs. Other advancements, such as smart particle packing and no-slip boundary condition in complex pore geometries, are also implemented for the construction and the simulation of the realistic shale pores from 3D nanometer-resolution stack images. Our code is validated for accuracy and compared against the CPU counterpart for speedup. In our benchmark tests, the code delivers nearly perfect strong scaling and weak scaling (with up to 512 million particles) on up to 512 K20X GPUs on Oak Ridge National Laboratory's (ORNL) Titan supercomputer. Moreover, a single-GPU benchmark on ORNL's SummitDev and IBM's AC922 suggests that the host-to-device NVLink can boost performance over PCIe by a remarkable 40\%. Lastly, we demonstrate, through a flow simulation in realistic shale pores, that the CPU counterpart requires 840 Power9 cores to rival the performance delivered by our package with four V100 GPUs on ORNL's Summit architecture. This simulation package enables quick-turnaround and high-throughput mesoscopic numerical simulations for investigating complex flow phenomena in nano- to micro-porous rocks with realistic pore geometries.

\vspace{1em}
\noindent {\it Keywords:} digital rock physics; shale; GPU; dissipative particle dynamics; multiphase flow

\section*{Program summary}

\noindent{\bf Program title:} \UMMDPD

\noindent{\bf Licensing provisions:} GNU General Public License 3

\noindent{\bf Programming language:} CUDA C/C++ with MPI and OpenMP

\noindent{\bf Nature of problem:} Particle-based simulation of multiphase flow and fluid-solid interaction in nano- to micro-scale pore networks of arbitrary pore geometries.

\noindent{\bf Solution method:} Fluid particles and solid wall particles are modeled with a many-body dissipative particle dynamics (mDPD) model -- a mesoscopic model for coarse-grained fluid and solid molecules. The pore surface wall boundary for arbitrary surface geometries is modeled with a no-slip boundary condition for fluid particles that prevents fluid particles from indefinitely penetrating in the walls. The time evolution of the system is integrated using the Velocity-Verlet algorithm. 

\noindent{\bf Restrictions:} The code is compatible with NVIDIA GPUs with compute capability 3.0 and above.

\noindent{\bf Unusual features:} The code is implemented on GPGPUs with significantly improved speed.

\section{Introduction}



Approximately $75\%$ of the sedimentary rocks on Earth are clastic nanoporous tight rocks, which are often referred to as shale. Shale contains most of the world's fossil energy sources (e.g. oil and natural gas). However, only a small fraction of the sources in shale can be recovered so far, in part due to the gaps of our knowledge in the relevant fundamental physics that ultimately control the dynamics of fluids in shale, which manifests extremely low permeability in the micro- to nano-Darcy range with average pore sizes from a few nanometers ($10^{-9}$ m) to a few micrometers ($10^{-6}$ m). Filling these knowledge gaps may help the development of more effective shale source recovery strategies. Most of the theories of fluid flow in geomaterials (and the predictive models built upon such theories) have been based on the concepts of classical continuum fluid dynamics and a rigid porous or fractured solid porous matrix, which assume ideal non-slip boundary conditions for fluid flow and transport \citep{bear1973dynamics}. Those concepts and models have proven adequate for developing the theories of single- and multi-phase flow in permeable porous media such as aquifers, soils, and conventional oil and gas reservoirs. Many pore-scale fluid flow models have been developed in either Eulerian or Lagrangian frame, based on the continuum computational fluid dynamics (CFD), e.g., the models based on lattice Boltzmann method (LBM) \citep{pan2001pore,pan2004lattice}, smoothed particle hydrodynamics (SPH) \citep{tartakovsky2005smoothed,tartakovsky2006pore}, and volume-of-fluid finite volume method (VOF-FVM) \citep{huang2005computer,huang2005modeling}. However, the behavior of fluids in nanoporous tight shale is very different, as the discreteness of molecules may impact flow and transport processes at higher scales, and the solid organic materials may play an important role as mechanical components, sorbents and sources of fluids. Besides, the large specific surface areas can make surface reactions and surface transport more profound. For example, in an ideal spherical pore of 100 nm diameter, about 6\% of the fluid is within a distance of 1 nm from the solid surface, whereas in a pore of 10 nm diameter, over 49\% of the fluid is within a distance of 1 nm, where the physical and chemical properties of the fluid can be significantly different from those of bulk fluids. A good understanding of large-scale flow and transport behaviors in shale requires robust and accurate multiscale computational models that can bridge the scale gaps between fluid molecular dynamics (MD) models and nanopore-scale fluid flow models.

\begin{figure}[ht]
\centering
\includegraphics[width=0.9\linewidth]{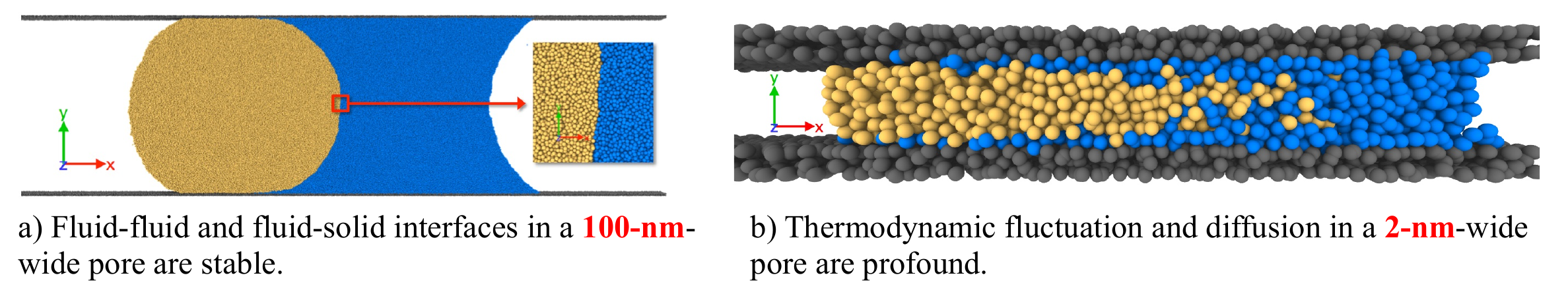}
\caption{Comparison of pore size effect on the continuum- and nano-scale fluid-fluid/solid interfaces in a slit-shape pore, as simulated by the mDPD model.}
\label{Fig:pore-size}
\end{figure}

Dissipative particle dynamics (DPD) constitutes a relatively new class of mesoscale models that can be used to simulate single- and multi-phase fluid flow \citep{tiwari2006dissipative,heldele2006micro,visser2006modelling,liu2006dissipative,liu2007dissipative-jcp,liu2007dissipative-wrr}. The DPD concept was originally introduced for microscopic hydrodynamics \citep{hoogerbrugge1992simulating} with its theoretical foundation based on statistical mechanics \citep{espanol1995statistical,marsh1998theoretical}. The various DPD models and their applications are summarized by \citet{moeendarbary2009dissipative} and \citet{liu2015dissipative}, respectively. In DPD, a system can be simulated with a set of interacting particles, where each particle represents a small cluster of molecules instead of a single one. The particle-particle interaction force in a DPD embodiment consists of a ``conservative'' (non-dissipative) component, a dissipative component that represents the effect of viscosity, and a thermal component that represents fluctuation. The distinction between DPD and SPH is the thermally driven fluctuations that are only detectable on microscopic scales, e.g. pores with sizes in the nanometer ranges. Conversely, DPD fluids can recover the continuum Navier-Stokes equations on large scales (scales much greater than the particle size) with the effect of thermal fluctuations to be negligible. Furthermore, DPD conserves mass and momentum, and also the energy provided with special treatment \citep{espanol1997dissipative,ripoll1998dissipative,avalos1999dynamic,li2014energy}), and allows much larger time steps than MD simulations. These features make DPD essentially a mesoscale method between the molecular and continuum hydrodynamic scales, and facilitates simulations of complex fluid systems with possible physical scales spanning a wide range. Recently, a so-called ``many-body'' DPD model \citep{warren2003vapor}, namely mDPD, has been found particularly suitable for multi-phase fluid systems, and thus has been applied for various multi-phase fluid simulation problems, including liquid-vapor interface, surface tension, and multi-component fluid flows in micro-scale channels \citep{pan2010single,chen2011many,ghoufi2011mesoscale,chen2013effective,chen2014many}. In particular, mDPD manifests a unique multiscale modeling capability that can model fluid-fluid/solid interfaces in pores at both continuum- and sub-continuum-scales, as demonstrated in \autoref{Fig:pore-size}.

Recently we developed an mDPD based nano to micro-scale pore flow model and applied it for multiphase flow simulations in source shale \citep{xia2017many}. In that model, realistic shale pore geometries are constructed based on 3D voxel data of shale core samples, which are generated from a focused ion beam scanning electron microscopy (FIB-SEM) digital rock imaging process \citep{goral2015pore} with voxel resolution at tens of nanometers or even a few nanometers. Each voxel contains local composition information that can be used to identify phase boundaries in shale, e.g. interfaces between inorganic and organic solid matrices, between inorganic solid matrix and pores, and between organic solid matrix and pores. The integration of FIB-SEM to nano-pore flow simulations is a big step forward as compared with the earlier methods that used either manufactured or analytically described pore geometries \citep{liu2007dissipative-wrr}. Furthermore, it is worth noting that though FIB-SEM has been adopted for analyzing shale samples for a while \citep{curtis2010structural,curtis2011transmission,curtis2012development,curtis2012microstructural}, most of the early flow simulation methods applied to shale were continuum CFD models (e.g. a finite element model by \citet{dewers2012three}), whose theoretical legitimacy yet remain to be fully verified for heterogeneous nanoporous media like shale. In comparison, the mesoscopic nature of mDPD (as shown in \autoref{Fig:pore-size}) makes the model a competent candidate for the nano- to micro-pore flow simulations in shale.

In order to use mDPD for predicting the critical material properties of shale micro core samples such as permeability and relative permeability, pore flow simulations must be conducted at meaningful space and time scales that may require simulations of a system with $10^8$-$10^9$ particles and $10^7$-$10^8$ timesteps. These simulations are computationally demanding and require significant computing resources. In early exercises we used the DPD package \citep{li2013three} in LAMMPS \citep{plimpton1995fast}. The package takes advantage of the parallel computing readiness of LAMMPS and delivers satisfying scalability for homogeneous porous systems. However, it is not the case for shale. Due to the highly non-uniform pore distributions in shale, load imbalance emerges as a result of non-uniform particle distributions and force calculations across the processing ranks and has been a serious bottleneck for the package to achieve desired scalability even with adaptive load balancing. Indeed, compared with the theoretical advances in multiphase DPD models, the development of efficient parallel strategies for those models is left behind, especially for heterogeneous porous systems at the appropriate physical scales. Efficient HPC strategies such as GPUs are highly encouraged. Because of the particular suitability of the general-purpose GPUs (GPGPUs) for MD and coarse-grained MD-like particle simulations, GPU computing has been widely adopted for mesoscale particle models such as SPH \citep{dominguez2013new,dominguez2013optimization,xiong2013gpu} and LBM \citep{januszewski2014sailfish,calore2016massively,tomczak2019new}. Some basic DPD models have been implemented in GPU accelerated packages such as HOOMD-blue \citep{glaser2015strong}, GROMACS \citep{abraham2015gromacs} and LAMMPS-GPU \citep{brown2011implementing}. The implementation of more sophisticated DPD models is recently described by \citet{tang2014accelerating} and \citet{blumers2017gpu}. Their GPU codes have demonstrated excellent strong- and weak-scalability for DPD simulations.

\begin{figure}[ht]
\centering
\includegraphics[width=\linewidth]{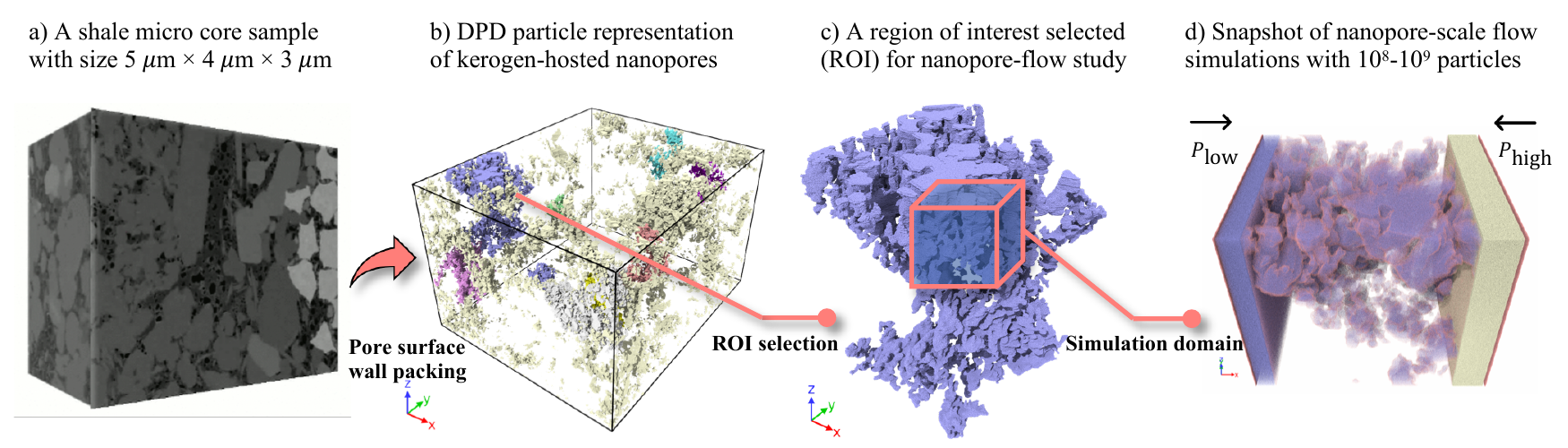}
\caption{Illustration of a production-level shale analysis workflow from nanometer-resolution digital rock imaging to GPU accelerated mDPD simulations of fluid flow in realistic nanopores in shale.}
\label{Fig:workflow}
\end{figure}

In this work, a generalized GPU-accelerated implementation of the mDPD based multiphase pore flow model with a solid wall boundary model for arbitrary pore geometries is developed to simulate flow dynamics in realistic source shale pores. The software features a tight integration of our earlier works including a mDPD pore flow model \citep{xia2017many}, an arbitrary-geometry wall boundary model \citep{li2018dissipative} and a GPU-accelerated DPD simulator \citep{tang2014accelerating,blumers2017gpu}, and delivers an efficient rock analysis throughput from digital rock imaging to pore flow simulations, as shown in \autoref{Fig:workflow}. With the new ability to model multiphase flow in arbitrary-shaped, nano- to micro-scale channels, the code package can be used to investigate the critical material properties of shale such as permeability and relative permeability with unprecedented time and length scales. Because a GPU can fit a workload comparable to many CPU codes, the use of GPUs can effectively reduce overhead in cross-rank/node communication. Consequentially the reduced rank-level parallelism is especially helpful for reducing load imbalance in mDPD flow simulations in non-uniform porous systems. For example, investing the same computing capacity, it requires a much smaller number of GPU cards than CPU cores, and hence much fewer ranks in GPU computing than CPU assuming one GPU card and one CPU core per rank. As a result, the use of GPUs would greatly reduce the number of domain decompositions in a non-uniform porous system, and thus is expected to improve load balance by substantially reducing cross-rank communication and latency in rank synchronization.

The rest of the paper is structured as follows. In \autoref{sec:model}, we briefly describe the mDPD model, a solid wall boundary model and surface wall particle packing for arbitrary geometries. In \autoref{sec:gpu-implementation}, we present the implementation and innovations of our program. In \autoref{sec:code-verification}, we validate the code with the verification problems. In \autoref{sec:benchmark}, we demonstrate the efficiency of our code by running benchmark cases for uniform and non-uniform nanoporous media. In \autoref{sec:capability-demonstration}, we further demonstrate the capability of the software with pore flow simulations in realistic shale nanopore networks. Lastly, we conclude the paper in \autoref{sec:summary}.

\section{Pore-scale fluid flow models \label{sec:model}}
 
\subsection{Many-body dissipative particle dynamics}

\begin{figure}[ht]
\centering
\subfloat[]{\includegraphics[width=0.3\linewidth]{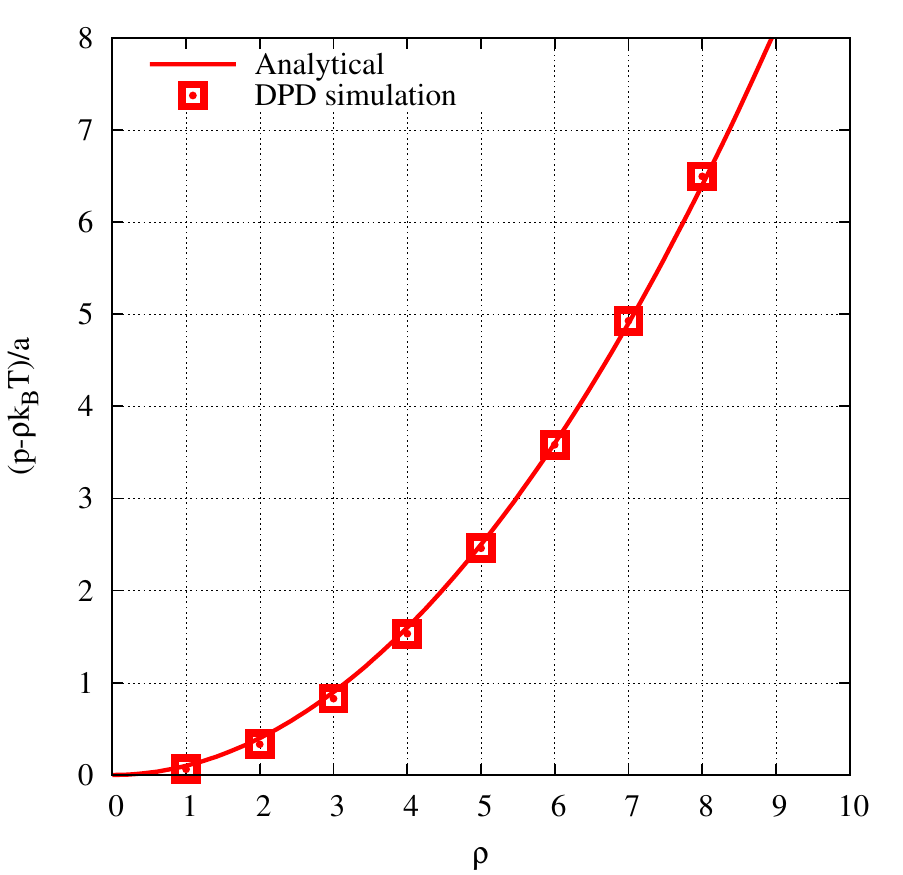}\label{fig:eos-dpd}}
\subfloat[]{\includegraphics[width=0.3\linewidth]{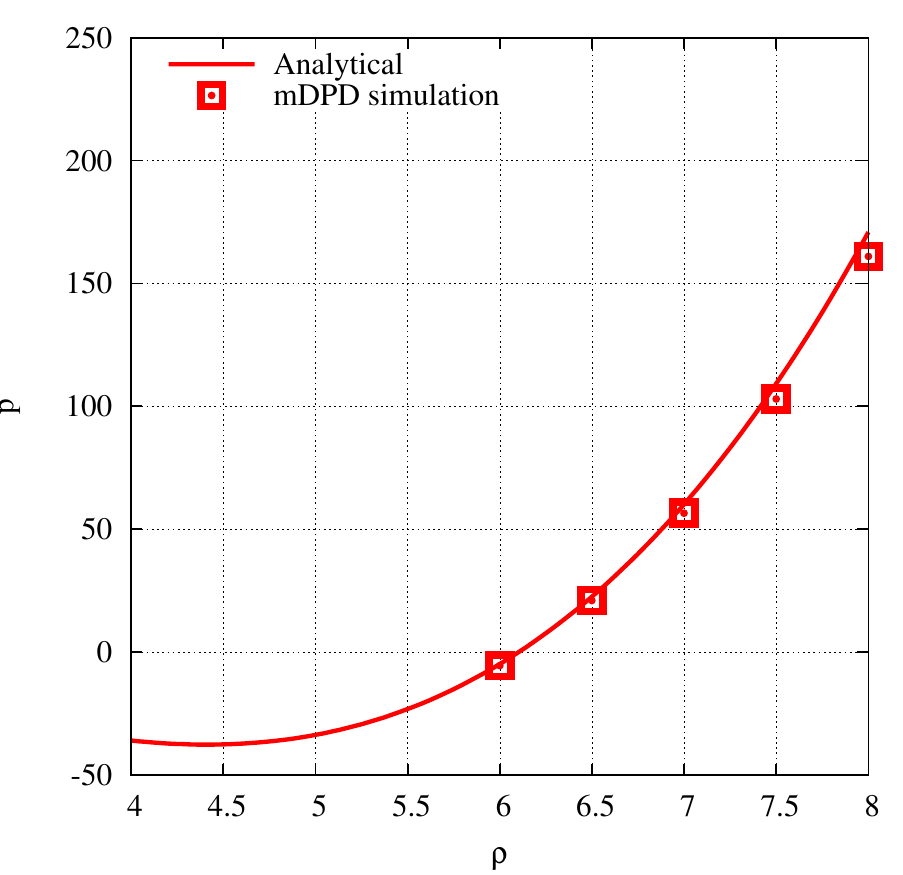}\label{fig:eos-mdpd}}
\caption{
Validation of the EOS: (a) $p=\rho k_B T + 0.1 a_{ij} r_C^4 \rho^2$
for the original DPD model with $k_B T = 1$, $\gamma=4.5$, $r_C=1$, and $a_{ij}=25$;
(b) $p = \rho k_B T + \alpha A_{ij} \rho^2 + 2\rho B_{ij} r_d^4 (\rho^3 - c\rho^2+d)$
for the mDPD model with $k_B T=1$, $\gamma=4.5$, $r_C=1$, $r_d=0.75$,
$A_{ij}=-40$, $B_{ij}=25$, $\alpha=0.101$, $c=4.16$, and $d=18$.
Pressure for each particle's number density $\rho$ is obtained
by averaging over $1000$ time steps after equilibrium, in a $10 \times 10 \times 10$ periodic box.
\label{fig:eos}}
\end{figure}

In a generic formulation, DPD particles interact via pairwise central forces, i.e. ${\bf F}_{ij} = {\bf F}_{ij}^{\rm R} + {\bf F}_{ij}^{\rm D} + {\bf F}_{ij}^{\rm C}$, where ${\bf F}_{ij}^{\rm R}$ represents a random force, ${\bf F}_{ij}^{\rm D}$ a dissipative force, and ${\bf F}_{ij}^{\rm C}$ a conservative force between particle $i$ and $j$, respectively. If ${\bf r}_i$ and ${\bf v}_i$ are used to denote the position and velocity of particle $i$, respectively, the random force ${\bf F}_{ij}^{\rm R}$ and the dissipative force ${\bf F}_{ij}^{\rm D}$ can be expressed as ${\bf F}_{ij}^{\rm R} = \sigma w^{\rm R} (r_{ij}) \xi_{ij} \hat{\bf r}_{ij}$ and ${\bf F}_{ij}^{\rm D} = -\gamma w^{\rm D}(r_{ij})(\hat{\bf r}_{ij}\cdot{\bf v}_{ij})\hat{\bf r}_{ij}$, where ${\bf r}_{ij} = {\bf r}_i - {\bf r}_j$, $r_{ij} = |{\bf r}_{ij}|$, $\hat{\bf r} = {\bf r}_{ij} / r_{ij}$ and ${\bf v}_{ij} = {\bf v}_i - {\bf v}_j$. These forces constitute a thermostat if the amplitude $\sigma$ of the random variable $\xi_{ij}$ and the viscous dissipation coefficient $\gamma$ satisfy a fluctuation-dissipation theorem: $\sigma^2 = 2 \gamma k_B T$ and $w^{\rm D}(r) = (w^{\rm R}(r_{ij} ))^2$, where $k_B T$ denotes the desired temperature in the unit of Boltzmann's constant $k_B$. In the original DPD model, the conservative force ${\bf F}_{ij}^{\rm C}$ is defined as ${\bf F}_{ij}^{\rm C} = a_{ij} w^{\rm C} (r_{ij}) \hat{\bf r}_{ij}$, where $a_{ij}$ denotes the magnitude of the force, and the weight function $w^{\rm C}(r)$ vanishes when the inter-particle distance $r$ is larger than a cutoff range $r_{\rm c}$. The ${\bf F}_{ij}^{\rm C}$ is usually derived from a soft and unspecific weight function $w^{\rm C}(r_{ij})$, thus allowing for a fairly large integration time step. Different weight functions describe different material properties. A common choice for $w^{\rm C}(r_{ij})$ is $w^{\rm C}(r_{ij}) = 1 - r_{ij} / r_{\rm c}$ and $w^{\rm R} = w^{\rm C}$. The standard velocity Verlet algorithm can be employed to integrate the resulting equations of motion in time. A quadratic equation of state (EOS) is obtained with respect to the average particle density $\rho$, as shown in Figure \ref{fig:eos-dpd}. However, the original DPD model is not sufficient to model multiphase fluid flow phenomena such as liquid-vapor interfaces, liquid-liquid interfaces and free capillary surfaces. A more complex EOS needs to be represented with the DPD model. To achieve this, a long-range attractive and short-range repulsive conservative force ${\bf F}^{\rm C}$ is required. The multiphase fluid flow model employed in the present work is the so-called many-body DPD method \citep{warren2003vapor}, namely mDPD. In mDPD, the ${\bf F}_{ij}^{\rm C}$ is augmented from the standard DPD method by density-dependent contributions, and the resulting model includes the van der Waals loop in the EOS, as shown in Figure \ref{fig:eos-mdpd}. In the mDPD model, the conservative force ${\bf F}_{ij}^{\rm C}$ is expressed as
\begin{equation}
{\bf F}_{ij}^{\rm C} =
A_{ij} w^{\rm C}(r_{ij})\hat{\bf r}_{ij} + B_{ij} (\bar{\rho}_i + \bar{\rho}_j) w_{\rm d}(r_{ij}) \hat{\bf r}_{ij}
\label{eq:mdpd-conservative-force}
\end{equation}
which consists of a long-range attractive part that is density-independent, and a short-range repulsive part that depends on a weighted average of the local particle density. The attractive component $A_{ij} w^{\rm C}(r_{ij})\hat{\bf r}_{ij}$ can be obtained by simply turning the sign of the original force parameter $a_{ij}$ (i.e., $A_{ij} < 0$, with a cutoff range $r_{\rm c} = 1$). The term $B_{ij} (\bar{\rho}_i + \bar{\rho}_j) w_{\rm d}(r_{ij})\hat{\bf r}_{ij}$ is a many-body repulsive component with $B_{ij} > 0$, and shorter cutoff $w_d (r_{ij}) = 1 - r/r_d$, where $r_d < r_C$. The averaged local density, $\bar{\rho}_i$ at the position of particle $i$ can be computed as $\bar{\rho}_i = \sum_{j \ne i}{w_{\rho} (r_{ij})}$, where the normalized weight function $w_{\rho}$ needs to satisfy $\int_{0}^{\infty}{4 \pi r^2 w_{\rho}(r)~dr} = 1$. For a three-dimensional computational domain, the $w_{\rho}$ is defined as $w_d(r) = \frac{15}{2\pi r_d^3}(1 - r/r_d)^2$.

\subsection{Solid wall conditions for arbitrary pore geometries \label{sec:solid-wall}}

Because of the soft particle-to-particle interaction in DPD models, fluid particles may penetrate through solid matrix given a fluid-solid interface. Such penetration is not physically possible and must be avoided. Early development of solid wall boundary models were focused on imposing rigorous macroscopic boundary conditions, e.g., a non-slip boundary condition at sharply defined impenetrable solid surfaces. The idea was from a strict mesoscopic interpretation of DPD models, where a single DPD fluid particle represents a cluster of fluid molecules on scales well above the atomistic levels \citep{groot1997dissipative}. To model a non-slip boundary, additional forces must be exerted on fluid particles at the vicinity of solid-fluid interfaces with model parameters carefully calibrated to avoid spurious behaviors such as artificial slip \citep{pivkin2005new}, temperature oscillation \citep{revenga1999boundary} and particle layering \citep{pivkin2006controlling}. To relax the strict non-slip requirement, \citet{henrich2007adhesive} proposed a boundary model, which imposes a weak external repelling force on fluid particles whenever they penetrate in solid matrix over a thin layer. However, most earlier boundary models are only suitable for solid surfaces that are either mostly flat, spherically curved, or at best analytically describable. A boundary model that can treat arbitrary pore geometries is required.

In this work, we adopt a new boundary model recently developed for DPD simulations involving arbitrarily complex geometries \citep{li2018dissipative}. For simulating pore flow in source rocks, this model enables construction of DPD systems of realistic nano- to micro-pore channels directly from loading the 3D stack images, so that the many intermediate steps from scanning electron microscopy (SEM) or transmission electron microscopy (TEM) images to the corresponding numerical models, i.e., surface mesh reconstruction, mesh smoothing and remeshing can be avoided. In particular, this boundary model computes a boundary volume fraction of fluid particles and allows the fluid particles to detect solid boundaries on-the-fly based on local particle configurations. As a result, with a negligible extra computational cost, the moving fluid particles become autonomous to find the pore surfaces and infer the wall penetration. A predictor-corrector algorithm is then applied to perfectly prevent the fluid particles from penetrating the pore surfaces. In addition, it is important to point out that by calculating and controlling the effective dissipative interactions between fluid and solid particles, the no-slip or partially-slip boundary condition are imposed on rough/curved pore surfaces with negligible density and temperature fluctuations in the vicinity of the solid boundary.

\subsection{Particle packing for pore surface geometries}

To construct bounding walls in DPD based fluid flow simulations, most researchers (e.g. \citet{meakin2007particle,chen2011many,li2018dissipative}) have followed a particle packing approach proposed in \citet{liu2007dissipative-wrr}. Using this packing approach, the whole simulation system will be first filled with DPD particles at a particle number density (e.g. $\rho_N = 8$) for solid matrix and then equilibrated. Next, particles located in defined flow regions will be deleted. To reduce cost, particles located in solid matrix but away from fluid-solid interfaces by over a specified distance will also be deleted, as those particles will have no interaction with fluid particles. The remaining particles are the so-called surface wall particles, whose coordinates will be saved and used as input data in wall-bounded flow simulations. This approach, though easy to use for relatively small systems, is however challenging for production-scale systems because of a temporary spike of computational and memory cost in the step of initial whole-system packing. The highest memory temporarily needed could be over 100 times higher than it may be eventually required, making it hardly affordable for most end users. For example, a shale micro core sample with a meaningful domain size might need billions of or even over a trillion particles to fill the system temporarily, but at last require no more than 1\% of them as surface wall particles because of the sample's low porosity.

\begin{figure}[ht]
\centering
\includegraphics[width=\linewidth]{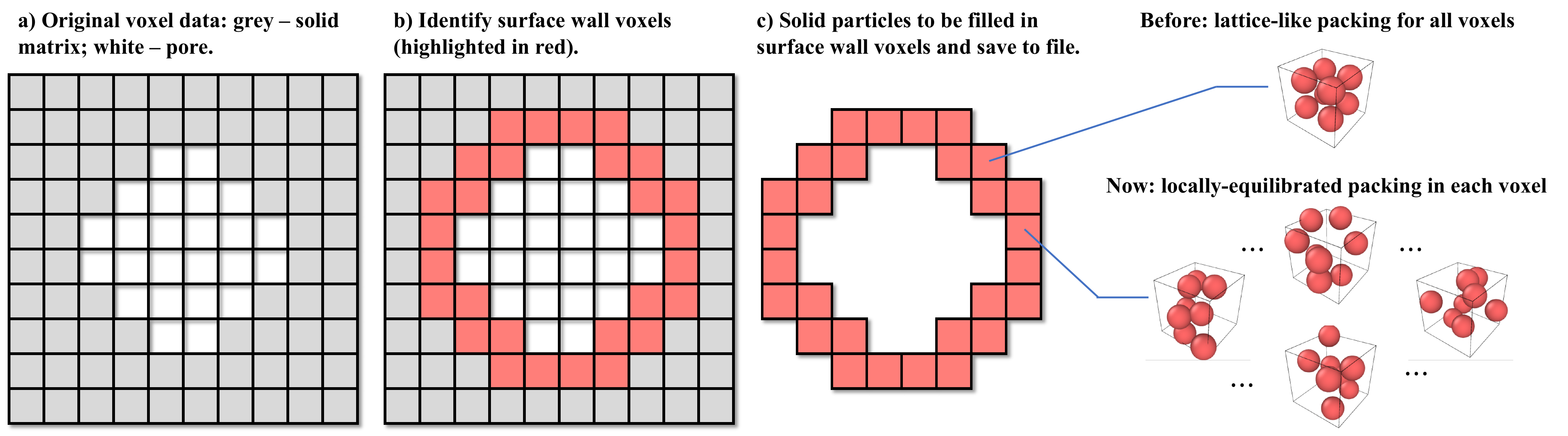}
\caption{Illustrations of a new additive particle packing process for constructing pore surface walls of porous systems based on 3D voxel data. To make it easy to understand, we use 2D pixels instead of 3D voxels in the display.}
\label{Fig:packing}
\end{figure}

For huge porous systems, to avoid the temporary but prohibitive computing and memory cost  incurred during the solid particle packing process, we introduce a new approach as an improved version of our early approach \citep{xia2017many}. Following our early version, a simulation system is determined based on voxel data of a shale micro core sample, in which each voxel records a numeric value for its local composition (e.g. pore, organic matter, or inorganic matter). An algorithm was developed to sweep through all the voxels to identify the so-called surface wall voxels, with the surface wall thickness equal to at least $r_{\rm c}$. In a second sweep, solid particles with a specified number density are created with a lattice-like distribution at locations corresponding to the surface wall voxels, and saved to data files for further use. Notice that the lattice-like packing of surface wall particles might cause undesired oscillations in fluid temperature in the vicinity of solid-fluid interfaces. Despite the known artifact, this approach had been probably the only affordable way for huge porous systems with arbitrary geometric complexity. To partially remedy the artifact, the present work proposes an improved particle insertion method. For each surface wall voxel, instead of employing the lattice-like packing, we use a locally equilibrated particle distribution that is randomly chosen from a database. The database is prepared in advance and is large enough for assembled pores to resemble sufficient randomness in pore surface roughness. \autoref{Fig:packing} is shown to illustrate this new packing method. Also notice that the idea of local equilibrium of the particles in each surface wall voxel makes the quality of packing closer to the one by \citet{liu2007dissipative-wrr}, but meantime would potentially give rise to non-equilibrium in particles across two neighbor surface wall voxels. Further improvement of affordable particle packing for pore surface walls in huge porous systems is an open area in DPD research.

\section{GPU implementation \label{sec:gpu-implementation}}

The present \UMMDPD package builds on \UMTWO \citep{blumers2017gpu}, which is a successor to the original fully GPU-accelerated \USERMESO package for DPD. \UMTWO expanded the capabilities of the package to simulate different flavors of DPD, as well as cellular dynamics. Although the new capabilities added in \UMMDPD only require the original \USERMESO \citep{tang2014accelerating} as base, we feel it more natural to name our software package \UMMDPD as a progression from \UMTWO.

\subsection{Core features}

The original \USERMESO \citep{tang2014accelerating} is a GPU-accelerated extension package to LAMMPS for DPD simulations. In the \USERMESO framework, all computations and host-device communications are handled by the extension package while I/O related tasks such as inter-rank communications are attended by LAMMPS. By offloading computations to GPUs, \USERMESO is able to achieves more than 20 times speedup for simple particle simulations \citep{tang2014accelerating}. The speedup over the CPU counterpart is made possible by technical innovations on, but not limited to, neighbor list constructions and particle reordering, which are intended to boost data locality and increases the chance of cache hit. Furthermore, data-layout is optimized for coalesced memory access. In LAMMPS, data are stored in an array-of-structure layout on host memory. To avoid strided access on device memory, data are stored in a structure-of-array layout. The conversion between the array-of-structure and structure-of-array layouts is carried out whenever data are transferred.

The notable innovative features of the original \USERMESO from which \UMMDPD has inherited include: 1) an atomics-free warp-synchronous neighbor list construction algorithm, 2) a two-level particle reordering scheme, which aligns with the cell list lattice boundaries for generating strictly monotonic neighbor list, 3) customized non-branching transcendental functions  (\texttt{sin}, \texttt{cos}, \texttt{pow}, \texttt{log}, \texttt{exp}, etc.), 4) overlapping calculation (e.g. force evaluation) with communication (e.g. particle exchange) to reduce latency, and 5) radix sort with GPU stream support.

\subsection{New capabilities}

To simulate complex single- and multi-phase fluid flow phenomena in realistic nano- to micro-porous geometries, a number of new features have been implemented in \UMMDPD.

A major contribution by \UMMDPD is the capability to run mDPD simulations. To recall the formulation in \autoref{eq:mdpd-conservative-force}, the many-body density $\rho$ that appears in the mDPD conservative force term is needed to calculate the repulsive part of the conservative form. For each particle, $\rho$ is computed immediately prior to the force computation. Then an inter-rank communication takes place to synchronize $\rho$ for the partition-ghost particles, as demonstrated in Algorithm \ref{alg:outline}.

Another important feature that has been implemented in \UMMDPD is the impenetrable wall boundary described in \autoref{sec:solid-wall} as a general solution to handle complex geometries in DPD simulations to treat pore surface walls of arbitrary geometric configuration. The main idea is to calculate the density of solid wall particles, $\phi$, within a fluid-particle's support, and then to add a correction force to the fluid particles to counter-react the artificial walls. Since $\phi$ is computed before the inter-rank communication, no synchronization is necessary as shown in Algorithm \ref{alg:outline}. 

\begin{algorithm}
\caption{An outline that depicts the calculation of many-body density $\rho$ and wall-particles density $\phi$ with reference to the Verlocity\-Verlet algorithm.}
\label{alg:outline}
\begin{algorithmic}[1]
\item Calculate $x(t + \delta t)$. 
\item $\rhd$ Calculate $\phi$ for all fluid-particles.
\item Inter-rank communication/particle migration.
\item $\rhd$ Calculate $\rho$ for all local particles.
\item $\rhd$ Synchronize $\rho$ for ghost particles.
\item Compute pair forces $f(t + \delta t)$.
\item Calculate $v( t + \delta t)$.
\end{algorithmic}
\end{algorithm}

\section{Code verification \label{sec:code-verification}}

In this section, we present two test problems to verify the implementation of the mDPD method and solid wall boundary condition in \UMMDPD. The numerical results calculated by \UMMDPD were verified with our CPU code, which is implemented based on the standard LAMMPS. Each problem underwent a comparative verification on two platforms: a workstation that has an Intel i7-8700K CPU and two NVIDIA TTIAN Xp GPUs, and a DGX-1 server that is equipped with two Intel Xeon E5-2698 v4 CPUs and eight NVIDIA Tesla V100 GPUs. 

\subsection{Liquid-vacuum interface \label{sec:liquid-vaccum-interface}}

In this problem, a simulation of water liquid-vacuum interface is presented with the objective to assess whether \UMMDPD accurately calculates properties of a specific type of fluid. The water density and surface tension calculated by \UMMDPD will be checked against its CPU counterpart. We followed the problem setup similar to \citet{ghoufi2011mesoscale}, but used a large cubic simulation domain bounded by $[-50r_{\rm c}, 50r_{\rm c}]$ in each direction with a periodic boundary condition. The simulation was initialized with a face-centered cubic ({\it fcc}) based particle allocation in the region of $x \in [-10r_{\rm c}, 10r_{\rm c}]$ and with a lattice spacing of $r_{\rm c}$ in each direction, which resulted in a total of $820,000$ particles in the system. The mDPD force interaction parameters $A_{ij} = -50$, $B_{ij} = 25$, $r_{\rm d} = 0.75r_{\rm c}$ and $\gamma = 12.4$ were used in order to match the water properties reported in \citet{ghoufi2010toward}. With those parameters, one DPD particle represents approximately a cluster of three water molecules (i.e., $N_m=3$), and the size of one DPD particle corresponds to about 90 \r{A}$^3$. Details of conversion from the reduced units to their corresponding physical values can be found in \citet{ghoufi2011mesoscale}.

\begin{figure}[ht]
\centering
\includegraphics[width=0.6\linewidth]{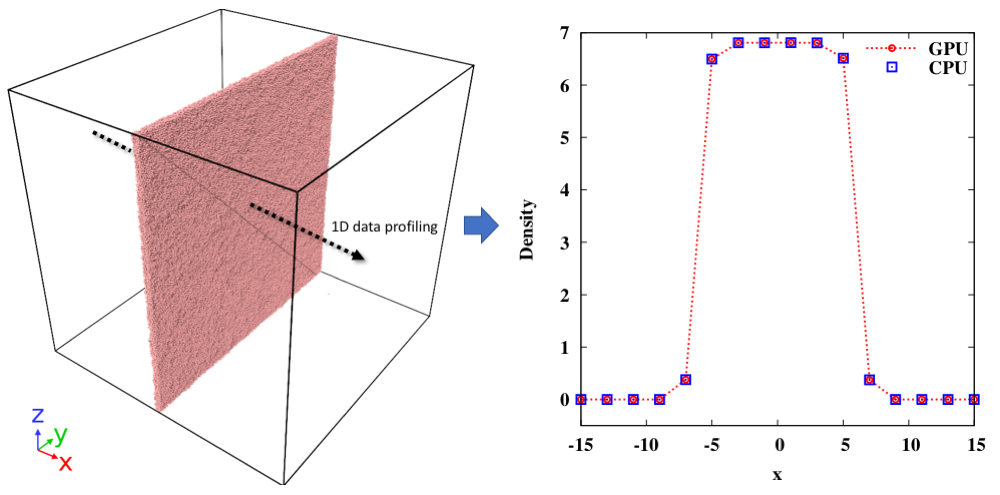}
\caption{A mDPD simulation of water liquid-vacuum interface: a snapshot of instantaneous particle distribution at equilibrium (left), and time-averaged density profile along the x direction (right).}
\label{Fig:water}
\end{figure}

In the simulation, a total of $5,000$ timesteps were first carried out to equilibrate the system. An instantaneous snapshot of of the equilibrated system is displayed on the left side of \autoref{Fig:water}, depicting a thin liquid slab formed by the particles. Another $5,000$ timesteps were then run to calculate the time-averaged properties. With a 1D bin size of $2r_{\rm c}$ along the x axis, a density profile calculated by \UMMDPD is compared with the one obtained by our CPU code on the right side of \autoref{Fig:water}. The density near $x = 0$ (center of the slab) is $6.88$ for both \UMMDPD and our CPU code, matching the value reported in \citet{ghoufi2011mesoscale}. Moreover, thanks to the simple shape of the liquid slab, the interfacial tension $\gamma_{\rm WV}$ between the water liquid and vacuum can be calculated by subtracting the mean tangential stresses $\sigma_{yy}$ and $\sigma_{zz}$ from the normal stress $\sigma_{xx}$: $\gamma_{\rm WV} = L_x \langle \sigma_{xx} - 1/2(\sigma_{yy} + \sigma_{zz}) \rangle$. The calculated $\gamma_{\rm WV}$ is $12.4$ for both \UMMDPD and its CPU counterpart, again matching the value reported in \citet{ghoufi2011mesoscale}. In addition, the values for water density and water-vacuum interfacial tension can be converted into the physical units with the equations: $r_{\rm c} = r_{\rm c}^* (\rho^* N_m V)^{1/3}$ [\r{A}], $\rho = \rho^* (N_m M)/(N_a r_{\rm c}^3 )$ $[{\rm kg} \cdot {\rm m} ^{-3}]$, and $\gamma = \gamma^* (k_B T)/(r_{\rm c}^2)$ $[{\rm N} \cdot {\rm m} ^{-1}]$, where the superscript * denote values in the reduced unit, $V$ is the volume of one water molecule ($30$ \r{A}), $M$ is the molar weight of a water molecule ($18$ ${\rm g} \cdot {\rm mol} ^{-1}$), $N_a$ is Avogadro's number, and $k_B$ is Boltzmann's constant, and $T$ is equal to $298$ K. Expressed in the converted physical units, the water density and liquid-vacuum interfacial tension are $\rho = 994$ ${\rm kg} \cdot {\rm m} ^{-3}$ and $\gamma = 70.6 \times 10^{-6}$ ${\rm N} \cdot {\rm m}^{-1}$, respectively, which agree well with the MD results \citep{ghoufi2011mesoscale}. Our result indicates that the implementation of the mDPD method in \UMMDPD achieves consistency with its CPU counterpart, and delivers accurate predictions of thermodynamic properties for fluids of interest.

\subsection{Static contact angle in a slit nano channel}

The second test problem is the simulation of static contact angles formed between a single fluid and its bounding solid walls in a slit nano channel, which demonstrates the flexibility of the mDPD model to characterize the wetting properties of fluids in the nano-scale pores. In the mDPD model, the particle interaction force between two types of materials such as solid and liquid can be modified by adjusting the attractive force parameter $A_{SL}$, the repulsive force parameter $B_{SL}$, and the repulsive force cutoff range $r_d$ in \autoref{eq:mdpd-conservative-force}, where the subscript ``S'' and ``L'' denote solid and liquid, respectively. In a controlled study of the dependence of liquid wetting behavior on certain mDPD parameters such as $A_{\rm SL}$, we selected three typical values for $A_{\rm SL}$ listed in \autoref{Tab:slit-pore-force-parameters}, while imposing constant values for the rest of the parameters, i.e. $B_{\rm SL} = 25$ and $r_{\rm c} = 1$ with a fixed relation between $r_{\rm d}$ and $r_{\rm c}$ as $r_{\rm d} = 0.75 r_{\rm c}$ for all particle interactions.
\begin{table}[ht]
\caption{Simulations of a single fluid in slid nano pore: specification of the attractive interaction parameters, $A_{\rm att}$.}
\scriptsize
\centering
\begin{tabularx}{\textwidth}{XXX}
\toprule
$A_{\rm att}$ & Solid & Lquid \\
\midrule
Solid &  -40 & -40 \\
Liquid & -40 & -35; -30; -20 \\
\bottomrule
\end{tabularx}
\label{Tab:slit-pore-force-parameters}
\end{table}

The simulation domain in this problem is bounded by $x \in [-30 r_{\rm c}, 30 r_{\rm c}]$, $y \in [-5r_{\rm c}, 5r_{\rm c}]$ and $z \in [-2.5r_{\rm c}, 2.5r_{\rm c}]$. A periodic boundary condition is prescribed in the x and z directions. The simulation consists of two steps. First, $3,500$ solid particles were initially placed in the two regions bounded by $y \in [-5r_{\rm c}, -4r_{\rm c}]$ and $[4r_{\rm c}, 5r_{\rm c}]$, respectively, with a random spatial distribution. These two regions were treated as two subsystems to allow the solid particles to undergo sufficient timesteps with the mDPD method to reach equilibrium. The locations of the solid particles were then fixed to represent the bounding walls of the slit pore for the rest of the simulations. The width of the slit pore (along the y direction) is $8r_{\rm c}$, corresponding to $8.616$ nm in the physical unit. Secondly, $4,000$ liquid particles were placed randomly in a region bounded by $x \in [-13r_{\rm c}, 13r_{\rm c}]$ and $z \in [-4r_{\rm c}, 4r_{\rm c}]$. The whole system was run for $4,000$ time steps to reach equilibrium using the mDPD model along with the solid wall condition. Finally, $10,000$ timesteps were run to obtain the time-averaged properties of interest. This simulation was performed three times with the three $A_{\rm SL}$ values, respectively.

\begin{figure}[ht]
\centering
\includegraphics[width=0.9\linewidth]{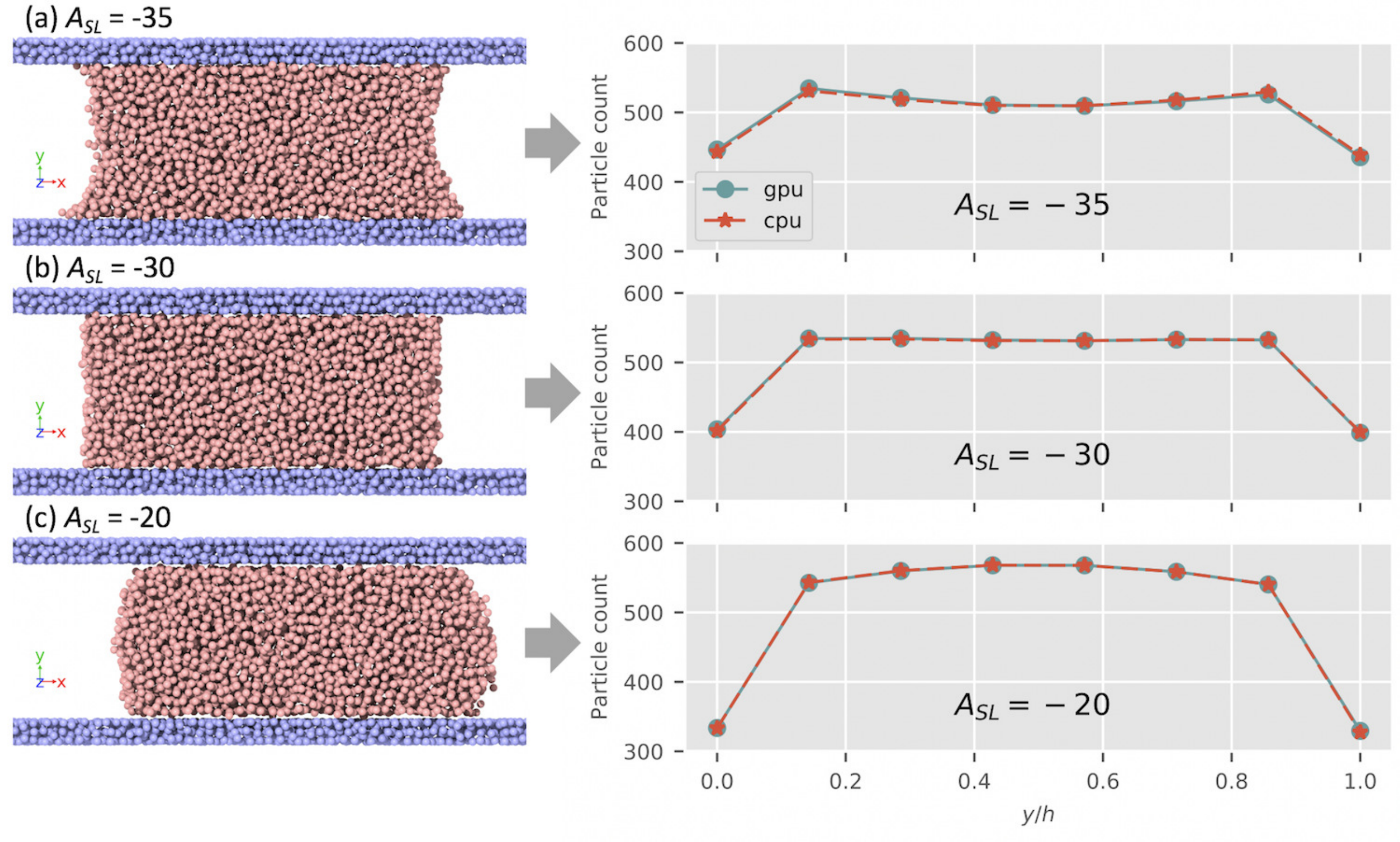}
\caption{Instantaneous particle distribution of a single liquid bounded by solid matter in a nanometer-scale slit pore, simulated by mDPD with different attractive force parameter $A_{\rm SL}$}
\label{Fig:vv-static-CA}
\end{figure}

The instantaneous snapshots of the particle distributions corresponding to the $A_{\rm SL}$ values are displayed one the left side of \autoref{Fig:vv-static-CA}, demonstrating the transition of the fluid wettability in the slit pore from wetting to non-wetting. Note that in the latter case, the fluid had shifted slightly away from its initial location due to the coupled effect of non-smooth wall surface and strong non-wettability of the fluid. To validate the consistency of \UMMDPD against its CPU counterpart, we plotted the profiles of the time-averaged fluid particle numbers versus the normalized pore width, and presented the GPU and CPU results on the right side of \autoref{Fig:vv-static-CA}. Eight bins were specified along the y direction, resulting in the eight data points in each profile. The GPU profiles agrees with their CPU references, indicating the numerical consistency. Furthermore, by dismissing the two near-wall points in those profiles, the curvatures of the profiles can be used to quantify the contact angles. For example, a higher $A_{\rm SL}$ such as $-35$ led to a partially wetting fluid with a contact angle smaller than $90^{\circ}$, whereas a lower $A_{\rm SL}$ such as $-20$ led in a partially non-wetting fluid with a contact angle larger than $90^{\circ}$. In the case of $A_{\rm SL} =  -30$, the profile is almost a straight line, depicting the critical state of contact angle around $90^{\circ}$. It is worth noting that a different choice in other parameters can result in a different dependency pattern of contact angle on $A_{\rm SL}$; for example, see a similar simulation in \citet{pan2010single}.

\section{Benchmark tests \label{sec:benchmark}}

In order to present a comprehensive performance benchmark, we tested \UMMDPD with simulations of fluid flows in both simple homogeneous and complex heterogeneous pore networks. HPC resources at Oak Ridge National Laboratory (ORNL), IBM and Idaho National Laboratory (INL) were used to perform the tests. We used the NVIDIA NVCC compiler with -O3 optimization to compile the code. The CPU counterpart, which has also been implemented based on the standard LAMMPS in this work, is compiled with the GCC compiler with -O3 optimization as well. We first benchmarked our package on a manufactured, homogeneous pore network, which serves to verify the code integrity and identify any intrinsic bottlenecks. We then quantified the performance of the code with a miniature version of a realistic pore-network. For both cases, the walltimes are compared with their respective CPU counterparts.  

\subsection{Fluid flow in homogeneous nanoporous media \label{sec:homogeneous-pore}}

\subsubsection{Problem description}

\begin{figure}[ht]
\centering
\includegraphics[width=0.75\linewidth]{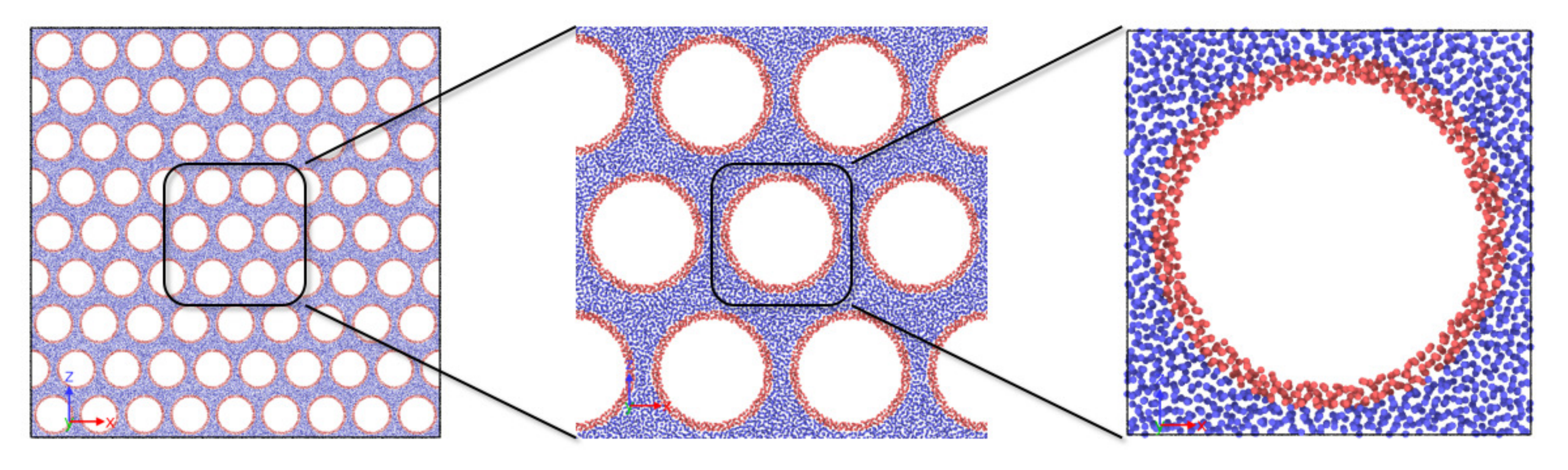}
\caption{Simulations of fluid flow in manufactured, homogeneous nanoporous media:
example of a porous domain consisting of $9^2$ square cells.}
\label{Fig:2d-pores}
\end{figure}

To showcase the scaling performance of \UMMDPD, body-force driven fluid flow was simulated in manufactured, homogeneous porous domains. Displayed in \autoref{Fig:2d-pores}, fluid flow in such a kind of domain is essentially two-dimensional, as the size of the domain in the y direction ($L_y$) is sufficiently small in comparison with the other two ($L_x$ and $L_z$). This domain is created based on a cell with $L_x = L_z = 16$ and $L_y=2$, as shown on the right side of \autoref{Fig:2d-pores}. We followed the procedure described in \citet{liu2007dissipative-wrr} to create such a cell, in which a ring-shape surface wall is constructed by $666$ equilibrated solid particles (red) with an outer radius of $7$ ($\approx 6.0$ nm) and an inner radius of $6$ ($\approx 5.1$ nm). Outside the ring, the space is filled with $1,296$ equilibrated fluid particles (blue). The cell is duplicated in the x and z directions (e.g. $23^2$, $33^2$ ... $65^2$ cells) to assemble a series of quasi-2D square domains, in which the even-numbered rows of cells are translated over a horizontal distance of $L_x / 2$ to finally form the domain for the flow simulations. For example, a domain consisting of $9^2$ cells is shown on the left side of \autoref{Fig:2d-pores}. These domains have a porosity of $0.4$, with the narrowest pore width to be $2$ ($\approx 1.7$ nm). The uniform pore distribution in this test minimizes load imbalance across the compute nodes. We thus consider it an appropriate problem to investigate the scalability of our code.

The mDPD force interaction parameters used in our previous work \citep{xia2017many} is adopted in this study. The attractive interaction parameters are listed in \autoref{Tab:2d-pores-force-parameters}, while the rest of the parameters used are $B_{\rm rep} = 25$, and $r_{\rm d} = 0.75 r_{\rm c}$ for all the particle-particle interactions. The particle number densities are $8$ and $6.2$ for the solid and fluid particles, respectively, ensuring that the pores are saturated at an adequate fluid pressure. An acceleration of $g_z = 0.02$ along the z direction is applied on the fluid particles to drive the flow. A periodic boundary condition is prescribed at all the three directions. A non-penetration boundary condition is prescribed at the solid particle wall surfaces. A timestep size of ${\rm dt} = 0.01$ is used. In each timing test, $10,000$ timesteps are run first to allow the domain to reach equilibrium under the influence of the fluid body force. The walltime is then measured for every $500$ timesteps, until four walltimes are obtained to calculate an average value.
\begin{table}[ht]
\caption{Simulations of fluid flow in manufactured, homogeneous nanoporous media:
specification of the mDPD particle-particle attractive interaction parameters, $A_{\rm att}$.}
\scriptsize
\centering
\begin{tabularx}{\textwidth}{XXX}
\toprule
$A_{\rm att}$ & Solid & Fluid \\
\midrule
Solid & --- & -40 \\
Fluid & -40 & -40 \\
\bottomrule
\end{tabularx}
\label{Tab:2d-pores-force-parameters}
\end{table}

\subsubsection{Benchmark results}

The scalability of our code is characterized with the strong- and weak-scaling performed on Titan at ORNL, Each Titan node is equipped with an AMD Opteron 6274 CPU, and a NVIDIA Tesla K20X GPU (Kepler architecture) with $2688$ CUDA cores and $6$ GB memory.

\begin{figure}[ht]
\centering
\includegraphics[width=0.6\linewidth]{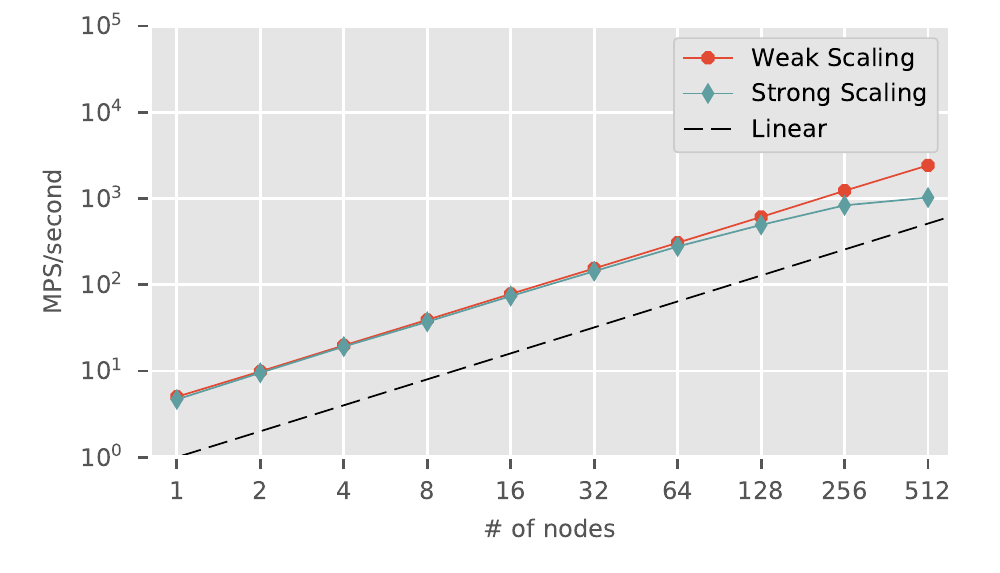}
\caption{Simulations of fluid flow in manufactured, homogeneous nanoporous media: the strong- and weak-scaling test results on the Titan supercomputer at Oak Ridge National Laboratory.}
\label{Fig:StrongWeak}
\end{figure}

For the strong-scaling, the test was carried out in a simulation system consisting of $33^2$ cells and a total of about $2.1$ million particles ($1.4$ million fluid particles and $0.7$ million solid particles). The system size was chosen to allow the memory of a single K20X GPU to accommodate the simulation. For the weak-scaling, the simulation system size was fixed at approximately $1$ million particles per node. The walltimes were obtained on systems consisting of $23^2$, $33^2$, $45^2$, $65^2$, $91^2$, $129^2$, $183^2$, $259^2$, $367^2$ and $519^2$ cells, respectively. To allow comparison across multiple platforms, the performance of our code was quantified with the metric ``million-particle-steps per second'', or MPS/second for short \citep{tang2014accelerating}. As shown in \autoref{Fig:StrongWeak}, our flow simulator scored a nearly perfect weak-scaling. On the other hand, the strong-scaling plot levelled off around $512$ nodes, when each node was loaded with approximately $4100$ particles.

\begin{figure}[ht]
\centering
\includegraphics[width=0.5\linewidth]{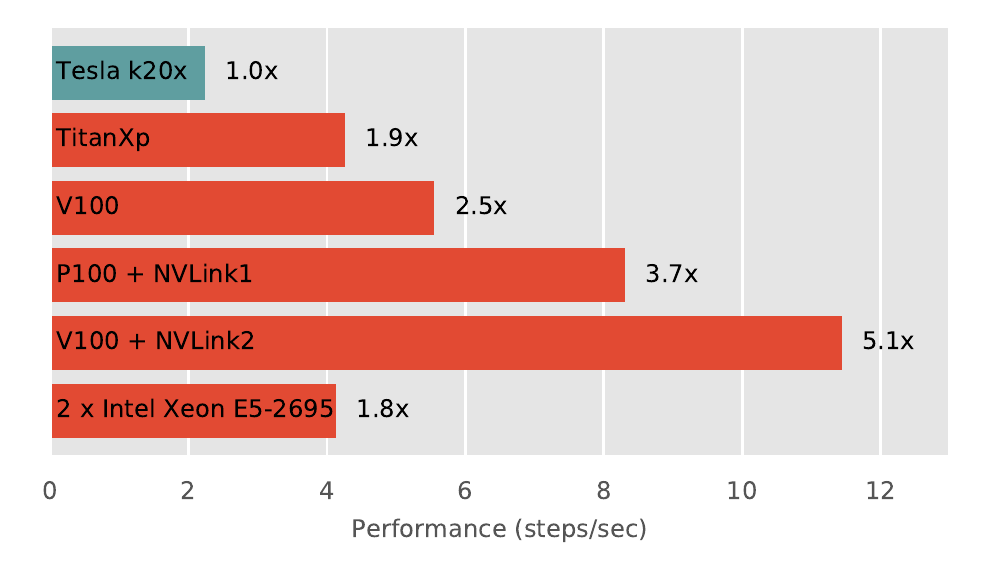}
\caption{Simulations of fluid flow in manufactured, homogeneous nanoporous media:
comparison of single-GPU performance on a number of latest GPUs.}
\label{fig:SingleGPU-performance}
\end{figure}

Besides the Tesla K20X, we benchmarked our code on a few more modern GPUs with advanced high-speed Host-to-Device interconnects to characterize the performance improvement brought by the latest hardware architectures. For clarity, the machines that have been tested are labelled and listed in \autoref{Tab:machines} with the detailed hardware specifications. Of particular note is the IBM AC922 node that is equipped with $42$ IBM Power9 cores and $6$ NVIDIA V100 GPUs with the NVLink2 interconnect: the same architecture configuration as ORNL's Summit supercomputer. To factor out Host-to-Host and/or node-to-node communication quality on different machines, we limited the comparative benchmark simulation running on one CPU core and one GPU on each machine. The walltime obtained on the Tesla K20X was used to serve as the baseline, while the performance of other machines was measured in terms of the relative speedup, as shown in \autoref{fig:SingleGPU-performance}.
\begin{table}[ht]
\caption{List of the hardware specifications for the labelled machines used in the benchmark test.}
\centering
\scriptsize
\begin{tabular}{llll}
\toprule
{\bf Label} (machine)                               & CPU                    & NVIDIA GPU  & Host-to-Device interconnect \\
\midrule
{\bf Tesla K20X} (ORNL Titan node)                  & AMD Opteron 6274       & Tesla K20X  & PCIe  \\
{\bf TITAN Xp} (desktop workstation)                & Intel i7-8700K         & TTIAN Xp    & PCIe   \\
{\bf V100} (NVIDIA DGX-1 at INL)                    & Intel Xeon E5-2698 v4  & Tesla V100  & PCIe   \\
{\bf P100 + NVLink1} (ORNL SummitDev node)          & IBM Power8             & Tesla P100  & NVLink1 \\
{\bf V100 + NVLink2} (IBM AC922 node)               & IBM Power9             & Tesla V100  & NVLink2 \\
{\bf 2 $\times$ Intel Xeon E5-2695} (INL HPC node)  & Intel Xeon E5-2695     & N/A         & N/A     \\
\bottomrule
\end{tabular}
\label{Tab:machines}
\end{table}

For the first, our test result has shown that the TITAN Xp (Pascal architecture, $3,840$ CUDA cores, $12$ GB memory), a top-tier consumer's model, produced nearly twice the performance of the Tesla K20X. Furthermore, our test result has shown that the Tesla V100 (Volta architecture, $5,120$ CUDA cores, $32$ GB memory) on DGX-1 can output $2.5\times$ the computing power of the Tesla K20X. On the other hand, because our code keeps the host and device memories separate for performance optimization, the overall performance depends heavily on the data transfer speed between the hosts and devices. In this regard, a remarkable finding is that the high-speed interconnects such as NVLink can dramatically shorten the walltime in our simulations. Together with the NVLink2 (the second-generation NVLink) on an IBM AC922 node, the V100 delivered an astonishing $5.1\times$ speedup over an ORNL Titan node. In other words, the NVLink2 is able to help double the performance of the V100 in our benchmark simulations. Lastly, to compare with the performance of a CPU-only implementation of our simulator, we benchmarked the CPU counterpart on an INL HPC node fully utilizing its $36$ cores (2 Intel Xeon E5-2695 v4 CPUs, 18 cores per CPU), and have found that it is equivalent to the TITAN Xp GPU in performance.

\begin{figure}[ht]
\centering
\includegraphics[width=0.5\linewidth]{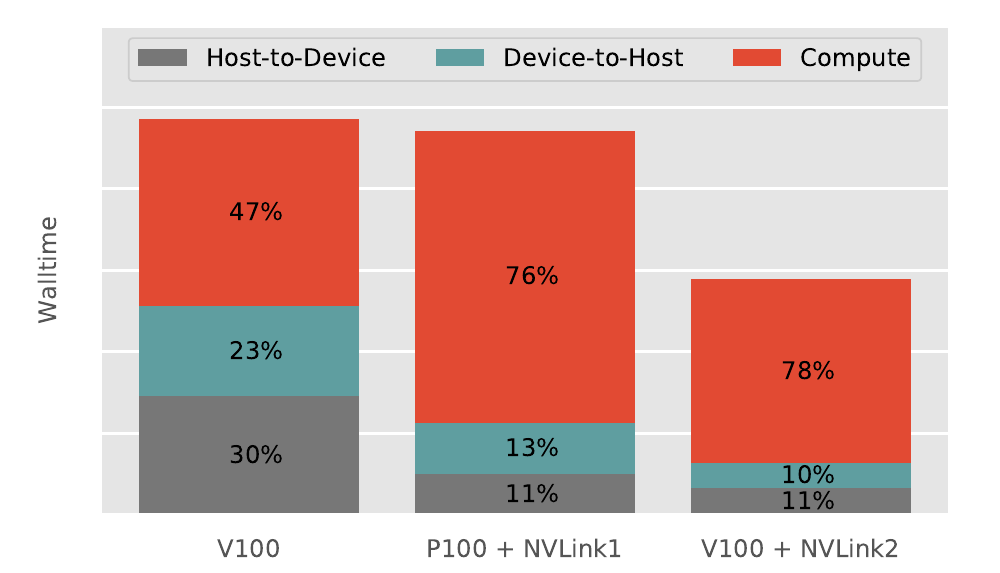}
\caption{Simulations of fluid flow in manufactured, homogeneous nanoporous media:
breakdown of walltime of a single-GPU simulation on GPU related tasks.}
\label{fig:SingleGPU-breakdown}
\end{figure}

With an interest to elaborate on the ramifications of the NVLink interconnect, we present a breakdown of the walltime on the GPU-related tasks in \autoref{fig:SingleGPU-breakdown}, e.g., Host-to-Device transfer, Device-to-Host transfer and kernel computation. For the Telsa V100 with the PCIe interconnect (DGX-1 node), the transfers together took up $53\%$ of the GPU related tasks (i.e., $30\%$ by Host-to-Device data transfer and $23\%$ by Device-to-Host data transfer). In comparison, when NVLink2 interconnected the host and the device, the transfers took up only $21\%$ while the walltime of kernel computations remains almost the same. In other words, NVLink2 has helped reduce the walltime of the GPU related tasks by about $40\%$ for our benchmark simulation. The same test was performed on SummitDev at ORNL (a tester cluster mimicking Summit), which has the Tesla P100 (Pascal architecture, $3,584$ CUDA cores, $16$ GB memory) with NVLink1 (the first-generation NVLink). Our result indicates that NVLink independently reduces considerable walltime that is sufficient to compensate for P100 when compared with its successor V100 without NVLink.

Above all, this benchmark problem has successfully demonstrated the excellent scalability of our code. Furthermore, the use of NVLink can drastically improve the efficiency of our code and provides performance boost to data-transfer intensive applications like our particle simulator. 

\subsection{Fluid flow in heterogeneous nanoporous media}

The objective of this problem is to assess and demonstrate the scaling performance of \UMMDPD for simulations of fluid flow in realistic heterogeneous nanopores, i.e., the shale kerogen-hosted pores. In this study, the construction of kerogen-hosted pores for pore-flow simulations was based on the nano-resolution stack images of a Vaca Muerta shale micro core sample, which refers to the geologic formation located at Neuqu$\acute{\rm e}$n Basin in Argentina \citep{badessich2016vaca}. The procedures for digital imaging of shale core samples and image post-processing for our pore-flow simulations are briefly described in Appendix \ref{appendix:shale-rock} for interested readers. Most hydrocarbons in shale are believed to be in kerogen-hosted pores before geotechnically processed. Massive hydrocarbon flow will not occur in kerogen with their natural low permeability \citep{teixeira2017microfracturing}. Permeability enhancement like hydraulic fracturing creates micro-cracks in shale and create linked paths for flow through connected pores spanning multiple scales (e.g. from nano- to micro-scale). Such structural evolution of organic-matter-hosted pores as well as the flow within is challenging to reproduce and measure in laboratory because of the required physical conditions \citep{panahi2019fluid}. Our benchmark test is thus focused on flow simulations in kerogen-hosted pores, in order to present an efficient pore-network flow simulation package for relevant research.

\subsubsection{Problem description}

\begin{figure}[ht]
\centering
\includegraphics[width=0.8\linewidth]{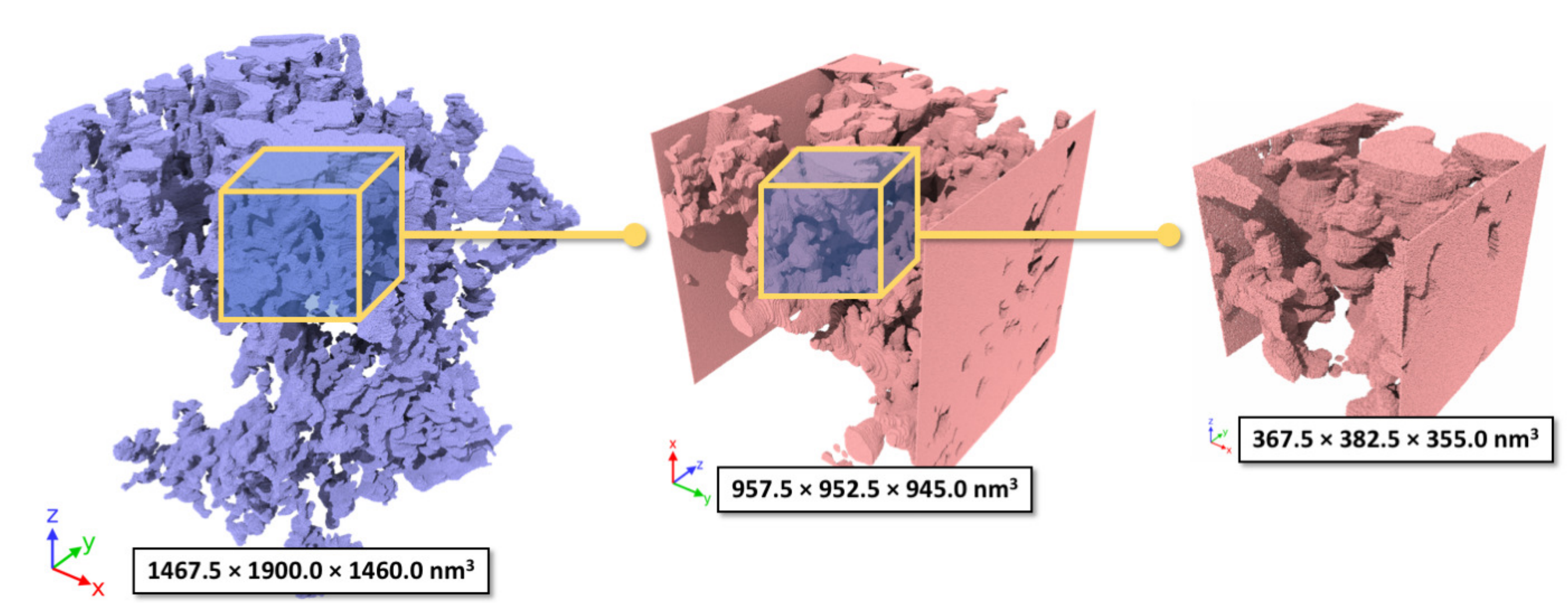}
\caption{Schematic for creation of block domains for flow simulations in organic-rich regions in a shale core sample.}
\label{fig:flooding-multiscale}
\end{figure}

For our benchmarking purpose, pore flow simulations in the entire core sample is not necessary. Instead, we focus on a large pore (labeled \#1) in \autoref{fig:shale-core} and introduce an example of how to set up a simulation domain for pore flow driven by bulk pressure gradient, as shown in \autoref{fig:flooding-multiscale}. In the first step, the \#1 pore is cropped to create a cubic block ($957.5 \times 952.5 \times 945.0$ nm$^3$), with two slabs perpendicular to a specified direction (e.g. x) added to the two ends of the block to allow fluid particles to move only inside the pore, as shown in \autoref{fig:flooding-multiscale} (middle). For flow simulation in this block, it is estimated to require over 200 million particles and 400 million timesteps. To allow the required memory to fit in a single V100 GPU for strong-scaling test, we cropped the block to a miniature version ($367.5 \times 382.5 \times 355.0$ nm$^3$), as shown in \autoref{fig:flooding-multiscale} (right).

\begin{figure}[ht]
\centering
\includegraphics[width=0.6\linewidth]{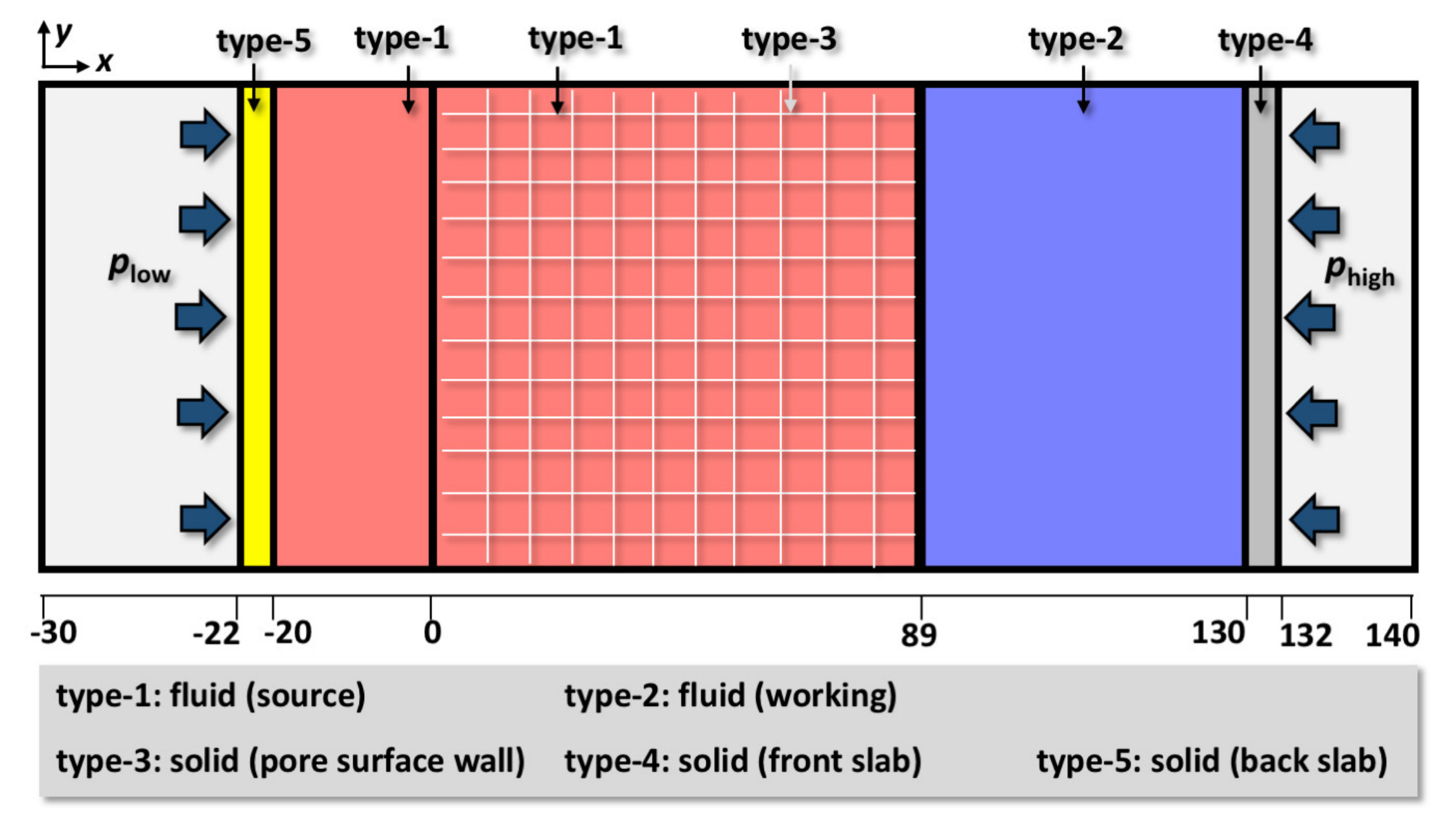}
\caption{Schematic for simulations of pressure gradient driven flooding in a block porous domain.}
\label{fig:mini-flooding-setup}
\end{figure}

The setup for our miniature version test is illustrated in \autoref{fig:mini-flooding-setup}, which is general enough for applying to a system of any size. The simulation box extents from -30 to 140 in x, 0 to 91 in y, and 0 to 88 in z, respectively. A reflection wall condition is prescribed at all the box boundaries to prevent fluid particles from accidentally fleeing, which though did not occur in our simulations. The simulation depicts a pressure gradient driven flooding through a porous block located at $x \in [0, 89]$. Five material types numbered from 1 to 5 are labeled for the particles. A total of 3,325,409 particles are created in the box, including 1,859,025 particles as type-1 fluid (source), 1,641,640 particles as type-2 fluid (working), 568,488 particles as type-3 solid (pore surface wall), and 128,128 particles for type-4 solid (front-pushing slab) and type-5 solid (back-pressure slab), respectively. Type-1 and 2 particles are assigned with the same mDPD model parameters as we consider single-phase flow in this study. Likewise, type-3, 4 and 5 particles represent solids of the same kind. The use of unique material types allows flexible change of model parameters.

\subsubsection{Benchmark results}

\begin{figure}[ht]
\centering
\subfloat[]{\includegraphics[width=0.32\linewidth]{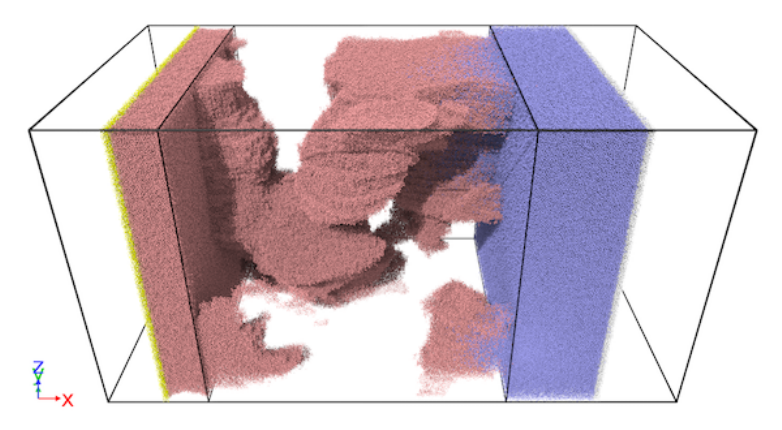}}
\subfloat[]{\includegraphics[width=0.32\linewidth]{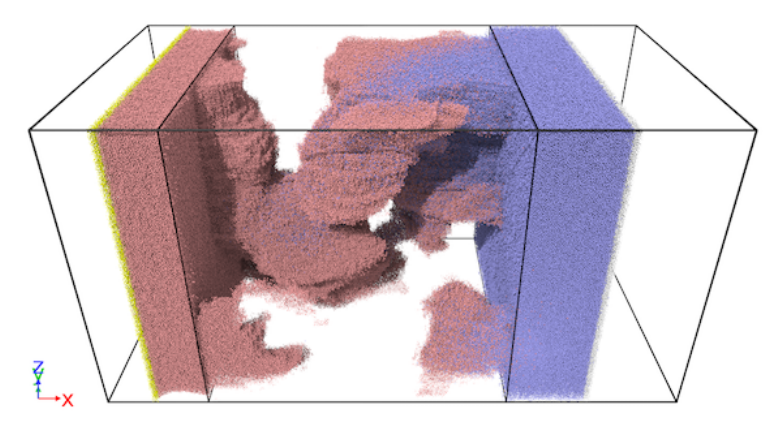}}
\subfloat[]{\includegraphics[width=0.32\linewidth]{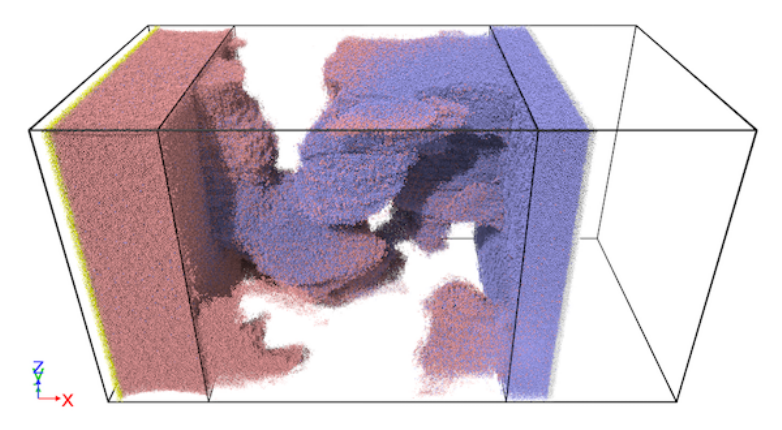}}
\caption{Miniature flooding test: a series of instantaneous snapshots for single-phase flooding in an organic-matter-hosted pore. The pore surface wall particles are not displayed, to allow fluid particles in the pore to be seen.}
\label{fig:mini-flooding-scene}
\end{figure}

The initial condition for the flooding simulation takes a few separate simulations to prepare. For the first, type-1 fluid particles are created to saturate the porous block (type-3). Extra type-1 fluid particles outside the block are pushed against the block by a slab (type-5) in order to sustain the hydraulic pressure in the pore. This setup mimicks hydrocarbons trapped in organic-matter-hosted pores. For the second, type-2 fluid particles are pushed against the block on the other side by a slab (type-4) with a higher external pressure. A virtual wall is placed at the boundary of the block ($x = 89$) to prevent type-2 fluid particles from entering the pore. At the beginning of the flooding simulation, the virtual wall is removed, and due to the bulk pressure difference between the two ends of the block, the type-2 fluid particles will be pushed into the pore gradually, while the type-1 fluid particles in the pore will be extracted. The mDPD model parameters and timestep size used in \autoref{sec:homogeneous-pore} are adopted here. A series of snapshots for the simulated flooding process are shown in \autoref{fig:mini-flooding-scene}, depicting the forced ejection of source fluid out of the pore.

\begin{figure}[ht]
\centering
\includegraphics[width=0.6\linewidth]{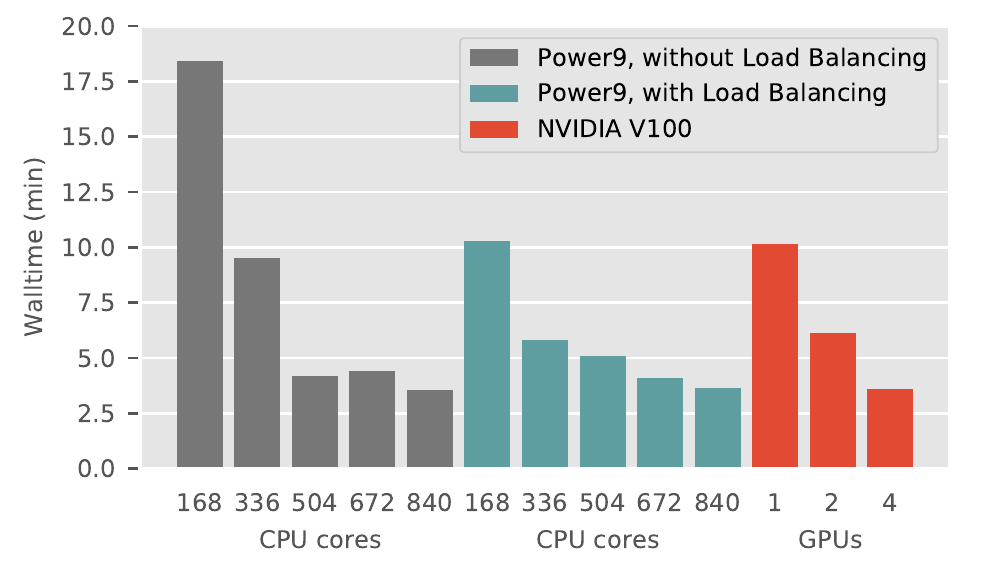}
\caption{A comparison of the walltime measured between \UMMDPD and its CPU counterpart for the miniature flooding simulations on the IBM AC922 cluster featuring Power9 CPUs and V100 GPUs with NVLink2.}
\label{fig:mini-walltime}
\end{figure}

To investigate the scalability of \UMMDPD on the flooding simulations in the realistic shale pore geometries, we carried out a set of strong-scaling tests using the Power9/V100 nodes on the IBM AC922 cluster. We chose the first 10,000 timesteps of the simulation for timing, during which the working fluid  rushes into the pore. Shown in \autoref{fig:mini-walltime}, the benchmark results indicate that the almost linear strong scaling obtained in \autoref{sec:homogeneous-pore} is no longer held true with the realistic nanopore geometries. This is because the fluid and solid particles are unevenly distributed in the simulation domain, unlike the uniform pore network described in \autoref{sec:homogeneous-pore}. When a simulation box is decomposed evenly based on the spatial dimensions, each subdomain has a distinctive particle composition tabulated in \autoref{Tab:miniflood}. As a result of the non-uniform particle distributions, the conventional spatial decomposition scheme does not offer a good strong scaling. Implementing a load balancing scheme such as the recursive coordinate bi-sectioning (RCB), the performance of the CPU code improved considerably, especially when fewer cores were used. For example, in our CPU timing with $168$ cores, the RCB cut the walltime almost in half. However, as more cores were engaged, the benefits of RCB subsided rapidly. This was observed in the CPU timing with $840$ cores, where the RCB failed to help reduce the walltime by a definitive amount. As for \UMMDPD, the conventional spatial decomposition is enforced in the current implementation. Furthermore, as a GPU can hold a much larger subdomain than a CPU core, the effect of load imbalance is much less pronounced. Hence despite the lack of load balancing schemes, \UMMDPD with $4$ V100 GPUs performed just as well as $840$ Power9 cores as seen in \autoref{fig:mini-walltime}, well demonstrating the superiority of GPU implementation for realistic complex geometries.

\begin{table}[ht]
\caption{Initial particle composition of each of the four subdomains. One subdomain is run on one GPU. The GPU with the heaviest workload is responsible for $38.7\%$ more particles than the one with the lightest workload.}
\centering
\scriptsize
\begin{tabularx}{\textwidth}{XXXXX}
\toprule
Subdomain & Fluids & Wall & Slabs & Total \\
\midrule
0  &   675,028 & 164,223 &  64,068 &   903,319 \\
1  &   830,701 & 182,201 &  64,060 & 1,076,962 \\
2  &   930,803 &  97,688 &  64,064 & 1,092,555 \\
3  & 1,064,133 & 124,376 &  64,064 & 1,252,573 \\
\bottomrule
\end{tabularx}
\label{Tab:miniflood}
\end{table}

\begin{figure}[ht]
\centering
\includegraphics[width=0.6\linewidth]{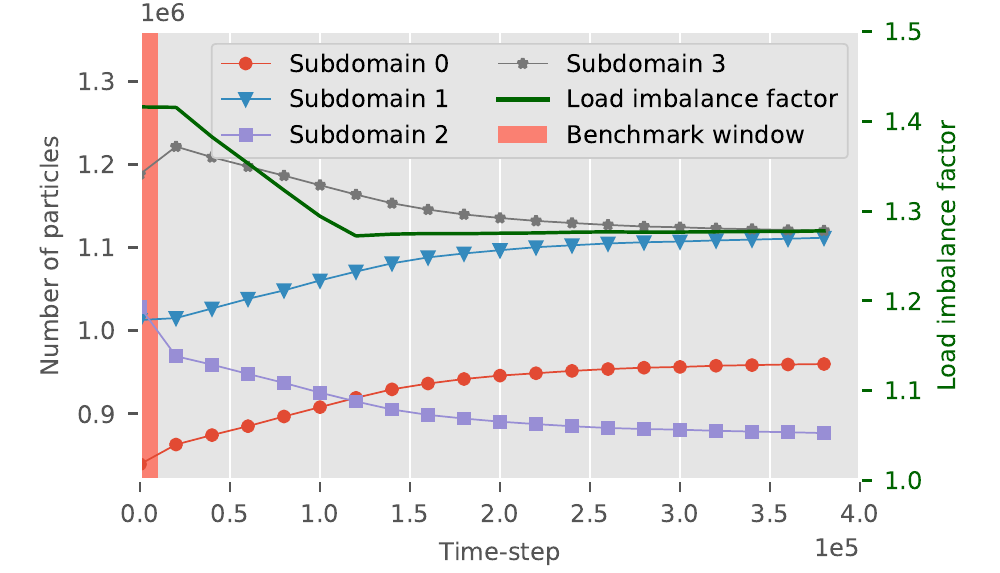}
\caption{Breakdown of the GPU workloads over the timesteps. Note that in the benchmark test between GPUs and CPUs, the walltimes were measured when the workload imbalance is the largest, indicated by the pink background.}
\label{fig:LoadBreakDown}
\end{figure}

To further illuminate the scalability challenge for the particle flow simulations in heterogeneous nanoporous geometries, we present a breakdown of the GPU workloads with four V100 GPUs and track the number of particles in each subdomain over the timesteps, as shown in \autoref{fig:LoadBreakDown}. Recall that the simulation box is evenly divided into four subdomains with one per GPU. We also plotted the load imbalance factor, which is defined as the ratio of the largest GPU workload to the smallest among the subdomains. The workload imbalance is the largest at the beginning of the simulatiton, when subdomain 3 contained approximately $25\%$ more particles than subdomain 0, corresponding to a load imbalance factor of $1.4$. As the working fluid rushed into the pore, the workloads became more even over time, and the factor descended to $1.28$ at most. Further investigation on the load balancing is not in the scope of this study. We intend to propose a general solution to control load imbalance on GPUs in a follow-up work.

\section{Capability demonstration \label{sec:capability-demonstration}}

\begin{figure}[ht]
\centering
\includegraphics[width=\textwidth]{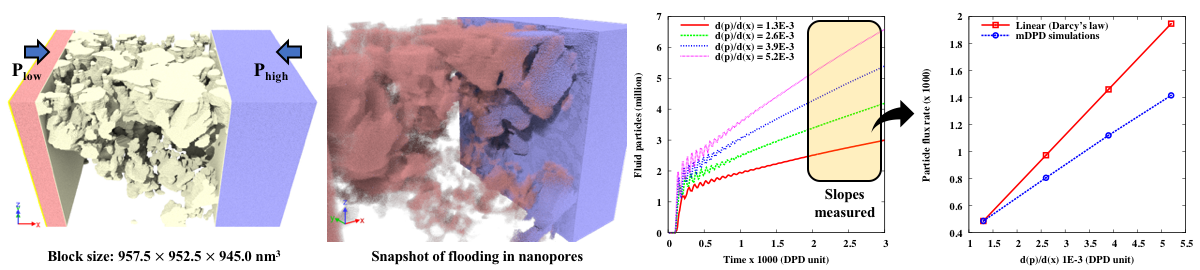}
\caption{Example of flooding simulations (about 240 million DPD particles) and permeability-fluid dependence characterization in a micro shale domain with realistic nanometer-resolution pore geometries.}
\label{fig:mid-flooding}
\end{figure}

Though it's a common understanding that the Darcy's law is no longer suitable for describing the flow and transport phenomena in nanoporous source shale rocks, so far no mature analytic formulation has been deduced experimentally to elaborate the source recovery processes in shale. Certain properties such as the permeability-fluid dependence (i.e. the correlation between the mass flow rate and bulk pressure gradient) are difficult to measure experimentally in the micro core samples. The \UMMDPD package presented in this work provides an alternative to characterize the fluid-permeability dependency with mesoscopic flow simulations in digitized nanometer-resolution realistic shale pore geometries. To demonstrate the versatility of our package, the micro block ($957.5 \times 952.5 \times 945.0$ nm$^3$) shown in the middle of \autoref{fig:flooding-multiscale} was used in the flooding simulations, with a brief depiction of the problem setup and a snapshot of the moving fluid particles on the left side of \autoref{fig:mid-flooding}. Again, for simplicity, we assumed single-phase flow by specifying the same model parameters for the working fluid (blue) and source fluid (red). The simulation box contained about 240 million particles. Four simulations corresponding to four successively increased bulk pressure gradients were performed. In each simulation, 3000 DPD time units were run to allow the mass flow rate to reach a stable status. A total of 2048 nodes on Titan at Oak Ridge National Laboratory were deployed for each simulation. The same simulation would take at least 15 times as long on the CPUs, deduced from our benchmark results.

Shown on the right side of \autoref{fig:mid-flooding}, the dependency of the flow rate on the bulk pressure gradient deviated from the Darcy's law, indicating a non-constant permeability in shale, in part because of their heterogeneous porosity distributions and the sub-continuum solid-fluid interactions in the nanopores. The simulation results coincide with the general observation from shale reservoir operations that the increased injection rate does not necessarily help increase the source recovery rate. However, as a case of capability demonstration, such limited simulations cannot provide all but a rough depiction of the complicated source recovery processes. An inclusive understanding can only be established with flow simulations based on a sufficiently large ensemble of shale core samples and a careful calibration of model parameters for specific types of fluids and solids.

\section{Summary \label{sec:summary}}

This work has presented a GPU-accelerated mesoscopic pore flow simulation package based on a many-body dissipative particle dynamics (mDPD) model to address the computational challenges in the numerical investigation of hydrocarbon flow in source shales. Leveraging mDPD's ability to model the sub-continuum and continuum flow phenomena, the complex flow dynamics and fluid-solid interactions in multiscale pore networks with pore sizes ranging from a few nanometers to a few micrometers can be resolved simultaneously. The effective use of GPUs enhances simulation performance significantly: almost linear scaling on up to 512 nodes is achieved in both our strong and weak scaling benchmarks, while further speedup is possible even beyond 1024 nodes. Besides, the use of the advanced device-to-host interconnects such as NVLink2 brings remarkable additional speedup over PCIe. Additional advances including the implementation of solid wall boundary conditions for mDPD flow in complex pore geometries and solid wall particle packing for huge systems have facilitated flow simulations in realistic shale nano pore networks that are constructed from 3D nanometer-resolution stack images. Furthermore, we have calculated the speedup over CPU counterpart through a realistic shale pore flow test: it requires 840 Power9 CPU cores to match the performance of 4 V100 GPUs on the Summit architecture. In summary, this package enables quick-turnaround and high-throughput mesoscopic numerical simulations for investigating complex flow phenomena in nano- to micro porous rocks with realistic pore geometries. We made our software freely available on GitHub, following the link \url{https://github.com/AnselGitAccount/USERMESO-2.0-mdpd}.

\section*{Acknowledgment}

The software development, validation and benchmark testing in this work is supported through the Idaho National Laboratory (INL) Laboratory Directed Research \& Development (LDRD) Program under the U.S. Department of Energy Idaho Operations Office Contract DE-AC07-05ID14517.
\\

\noindent The weak- and strong-scaling benchmarks and simulations for capability demonstration were primarily performed at Oak Ridge Leadership Computing Facility (OLCF) through the OLCF Director's Discretion Program under project GEO124, which is supported by the Office of Science of the U.S. Department of Energy under Contract DE-AC05-00OR22725.
\\

\noindent The benchmark testing also used resources in the High Performance Computing Center at INL, which is supported by the Office of Nuclear Energy of the U.S. Department of Energy and the Nuclear Science User Facilities under Contract No. DE-AC07-05ID14517.
\\

\noindent The numerical investigation of permeability-fluid dependence in shale kerogen-hosted nanopores was supported as part of the EFRC-MUSE, an Energy Frontier Research Center funded by the U.S. Department of Energy, Office of Science, Basic Energy Sciences under Award No. DE-SC0019285.

\bibliographystyle{plainnat}

\begin{appendices}
\section{Digital imaging and post-processing of shale core samples \label{appendix:shale-rock}}

\begin{figure}[ht]
\centering
\includegraphics[width=0.95\linewidth]{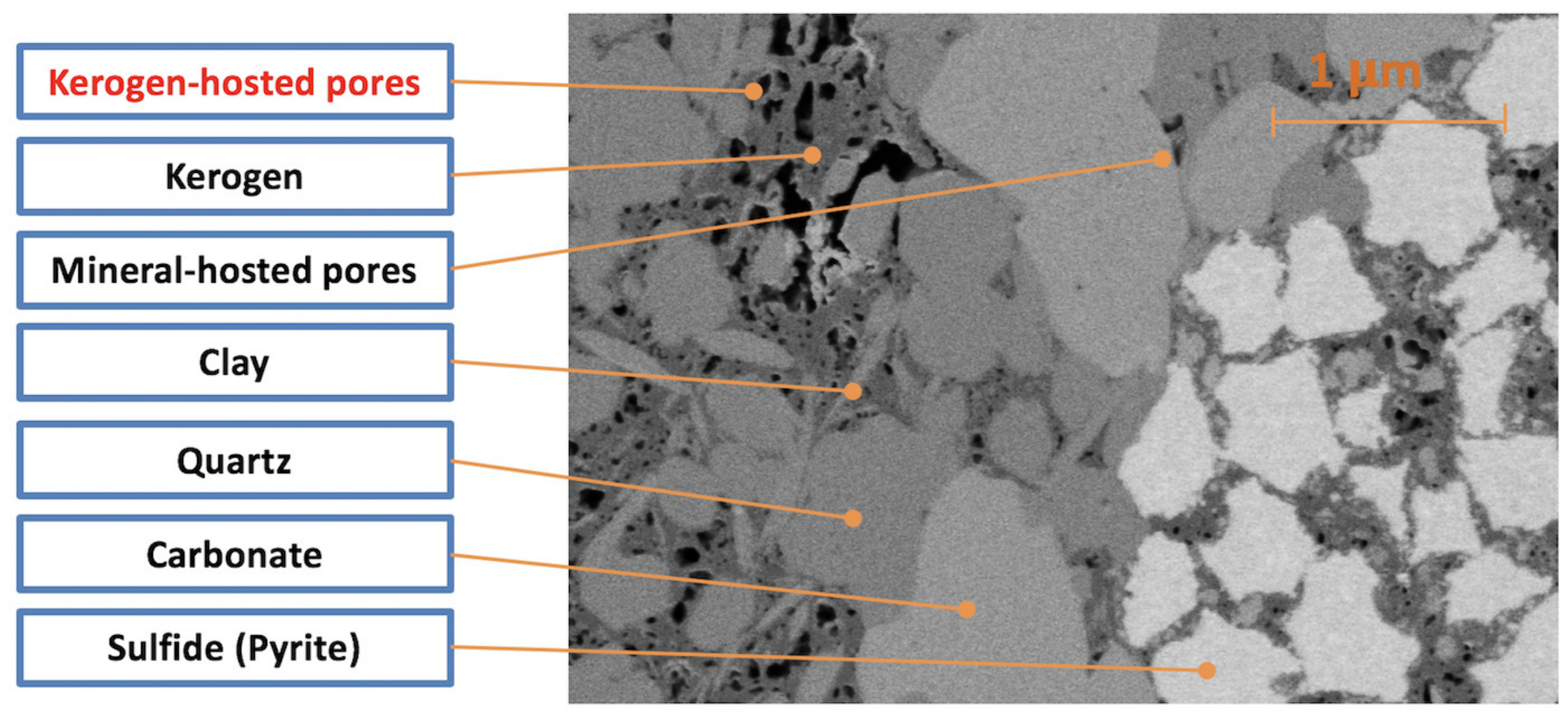}
\caption{A glance at the Vaca Muerta shale constituents in a raw digital image obtained by a FIB-SEM process. The resolution of this image is $2.5 \times 2.5$ nm$^2$ per pixel.}
\label{fig:shale-constituents}
\end{figure}

The Vaca Muerta shale micro core sample referred to in this work underwent a FIB-SEM process, which resulted in a stack of raw images with $2.5 \times 2.5$ nm$^2$ pixel resolution in each image and $5$ nm interval in scanning direction. \autoref{fig:shale-constituents} displays one of such raw images to illustrate the complex constituents in the sample. In a simplistic manner, we categorized the shale constituents in four phases: 1) inorganic matters, 2) inorganic-matter-hosted pores, 3) organic matters, and 4) organic-matter-hosted pores (i.e. kerogen-hosted pores). The raw images were not readily usable to pore-flow simulations because they could contain digital noises that should be filtered out first.

\begin{figure}[ht]
\centering
\includegraphics[width=0.95\linewidth]{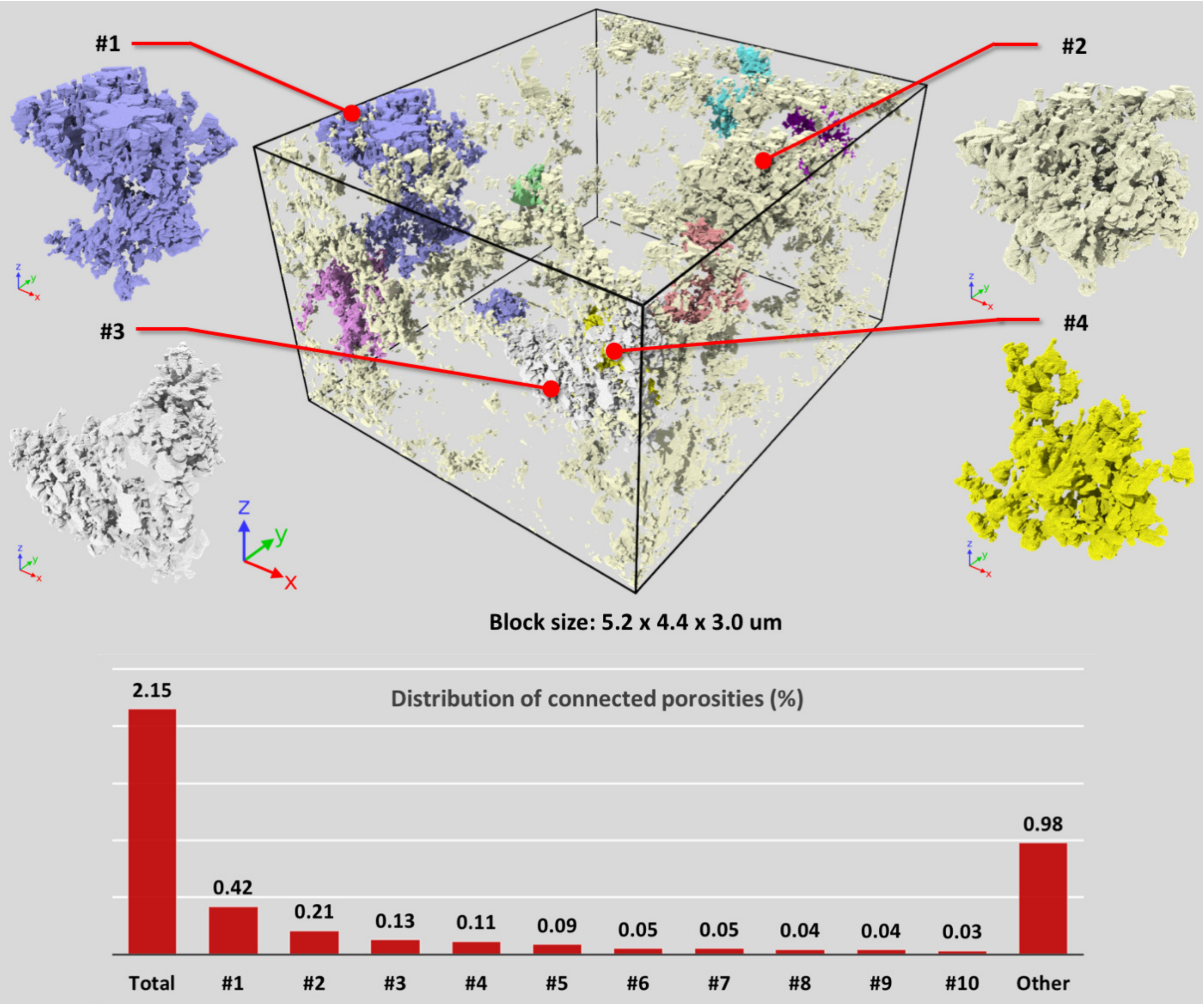}
\caption{Visualization of kerogen-hosted pores in a block region, with the ten largest pores rendered in unique colors and the top four largest pores labeled with \#1, \#2, \#3 and \#4. Other smaller and isolated pores are colored in light yellow. Bottom: distribution of the connected porosities (\%).}
\label{fig:shale-core}
\end{figure}

The raw images were post-processed with the Dragonfly image processing toolkit. The processed images were used for the preparation of DPD-based pore flow simulations. A block region of interest that contains an abundance of kerogen-hosted pores was found in our micro core sample and selected for preparation of the pore-flow simulations reported in this work. This block region has a size of width = 5,232.50 nm in width, height = 4,400 nm, and depth = 3,030 nm, and is visualized in \autoref{fig:shale-core}, where the pore networks are represented with pore surface wall particles generated with the image-to-particle workflow described in \autoref{sec:solid-wall}. In this block region, the ten largest pores that have no connectivity with others are each rendered with a unique color, and the rest of smaller isolated pores are colored in light yellow. The distribution of kerogen-hosted porosities in this block region is also reported in \autoref{fig:shale-core}, demonstrating the low-porosity feature of kerogen in shale as well as the discreteness of the pores.

\end{appendices}

\end{document}